\documentclass[12pt,oneside, a4paper]{article}

\pdfoutput=1  
 
\ifx\pdfoutput\undefined
\usepackage[dvips,bookmarks=false]{hyperref}	
\else
\usepackage{hyperref}	
\fi
\hypersetup{colorlinks,bookmarksopen,bookmarksnumbered,citecolor=blue,
linkcolor=black,pdfstartview=FitH,urlcolor=blue}

\usepackage{graphicx}
\usepackage{subfigure}
\usepackage{amssymb}
\usepackage{cite}
\usepackage{bm}
\usepackage{amsmath,amsthm}
\usepackage{cleveref}
\usepackage{siunitx}
\usepackage{tikz-feynman}
\usepackage{slashed}
\numberwithin{equation}{section}

\newcommand{\capdef}{}
\newcommand{\mycaption}[2][\capdef]{\renewcommand{\capdef}{#2}%
       \caption[#1]{{\footnotesize #2}}}

\begin{document}

\begin{titlepage}

\begin{center}

\vspace*{2cm}
        {\Large\bf Decoherence effects in reactor and Gallium neutrino oscillation experiments -- a QFT approach}
\vspace{1cm}

\renewcommand{\thefootnote}{\fnsymbol{footnote}}
{\bf Raphael Krueger}\footnote[1]{raphael.krueger@ruhr-uni-bochum.de},
{\bf Thomas Schwetz}\footnote[2]{schwetz@kit.edu}
\vspace{5mm}

{\it%
  {Institut f\"ur Astroteilchenphysik, Karlsruher Institut f\"ur Technologie (KIT),\\ 76021 Karlsruhe, Germany}}

\today
  
\vspace{8mm} 

\abstract{We adopt the quantum field theoretical method to calculate the amplitude and event rate for a neutrino oscillation experiment, considering neutrino production, propagation and detection as a single process. This method allows to take into account decoherence effects in the transition amplitude induced by the quantum mechanical uncertainties of all particles involved in the process. We extend the method to include coherence loss due to interactions with the environment, similar to collisional line broadening. In addition to generic decoherence induced at the amplitude level, the formalism allows to include, in a straightforward way, additional damping effects related to phase-space integrals over momenta of unobserved particles as well as other classical averaging effects. We apply this method to neutrino oscillation searches at reactor and Gallium experiments and confirm that quantum decoherence is many orders of magnitudes smaller than classical averaging effects and therefore unobservable. The method used here can be applied with minimal modifications also to other types of oscillation experiments, e.g., accelerator based beam experiments.}

\end{center}

\end{titlepage}

\renewcommand{\thefootnote}{\arabic{footnote}}
\setcounter{footnote}{0}

\tableofcontents

\section{Introduction}

Conceptual questions related to quantum decoherence effects in neutrino oscillations are the topic of ongoing discussions in the literature since several decades. For early papers on the topic see \cite{Nussinov:1976uw,Kayser:1981ye,Giunti:1991ca,Rich:1993wu,Kiers:1995zj,Grimus:1996av,Giunti:1997sk,Giunti:1997wq,Stodolsky:1998tc,Grimus:1998uh,Cardall:1999ze}, some examples of further investigations are e.g., 
\cite{Beuthe:2001rc,Akhmedov:2009rb,Akhmedov:2010ms,Hernandez:2011rs,Akhmedov:2012uu,Jones:2014sfa}.
Recently, this discussion received further attention in the context of short-baseline reactor
\cite{Ko:2016owz,Alekseev:2018efk,PROSPECT:2020sxr,STEREO:2022nzk}
and radioactive source Gallium experiments
\cite{Kaether:2010ag,Abdurashitov:2005tb,Barinov:2021asz,Barinov:2022wfh}
searching for sterile neutrino oscillations \cite{Arguelles:2022bvt,Akhmedov:2022bjs,Jones:2022cvh,Akhmedov:2022mal,Hardin:2022muu}. These papers discuss the question whether quantum decoherence could help to reduce tension in the data \cite{Arguelles:2022bvt,Hardin:2022muu} which arises in standard sterile neutrino explanations, see e.g., \cite{Dentler:2017tkw,Berryman:2021yan,Giunti:2021kab,Giunti:2022btk}.
Possible decoherence effects in the upcoming high-precision JUNO reactor experiment have been discussed in Refs.~\cite{deGouvea:2020hfl,JUNO:2021ydg,Marzec:2022mcz}.

In this paper we contribute to this discussion by adopting the quantum field theoretical (QFT) approach to neutrino oscillations \cite{Rich:1993wu,Giunti:1993se,Grimus:1996av,Kiers:1997pe,Beuthe:2001rc}. In this approach the combined process of neutrino production, propagation, and detection is considered as a single process whose amplitude is calculated by usual $S$-matrix methods, adapted to the situation of macroscopically separated production and detection regions. The neutrino is treated as an internal line and coherence properties of the flavour transition are completely determined by the localization of the external particles at the source and at the detector.

We will follow largely the methods discussed by Beuthe in the review article Ref.~\cite{Beuthe:2001rc}, with some modifications relevant to the experimental situations of interest to us. In particular, we generalize the formalism to take into account decaying particles (see also \cite{Grimus:1999ra,Akhmedov:2010ms,Grimus:2019hlq,grimus}) as well as coherence loss due to interactions of external particles with the environment. We focus on quantities which are actually observed in experiments, for instance the energy of the positron produced by the inverse beta-decay reaction in the detectors of reactor neutrino experiments. The formalism naturally allows a distinction of ``quantum decoherence'' (at the amplitude level) compared to ``classical averaging'' (at the probability level), which are, however, indistinguishable observationally. Our work is complementary to the recent paper by Akhmedov \& Smirnov~\cite{Akhmedov:2022bjs}, who base their argumentation on the neutrino wave packet approach, reaching very similar conclusions as we do. A somewhat different approach has been pursued by Jones, Marzec \& Spitz~\cite{Jones:2022hme} whose results for the decoherence parameters differ quantitatively from ours.

We provide here some guideline on how to read this paper. In \cref{sec:single-vertex} we introduce the notation by discussing external wave packets for single vertex processes, such as scattering or decay. Our ansatz of how to include decoherence effects due to interactions with the environment as well as decaying particles is introduced in \cref{sec:interact}. In \cref{sec:oscAmpl} we sketch the QFT calculations for the oscillation amplitudes. A reader mainly interested in the final result may, after a look at the central expression in \cref{eq:Asq2}, skip directly to \cref{sec:decoh-discussion}, which contains a brief discussion of the decoherence effects, and \cref{sec:classical}, where we comment on classical averaging and stress the equivalence of quantum and classical decoherence. \Cref{sec:numerics} contains the numerical estimates for reactor and Gallium source experiments, where in \cref{sec:localisation} we give some details of how we estimate the localizations of all involved particles. Readers interested only in the main numerical results may proceed directly to \cref{sec:eff-spread}, where we discuss the effective localization and energy spreads and show that they are many orders of magnitude below the observable level. We summarize our findings in \cref{sec:conclusions}. Supplementary material is provided in \cref{app:int,app:rates,app:phase-space}. In particular, \cref{app:rates} contains a discussion of how to derive standard expressions for scattering cross sections and decay rates within our formalism, and in \cref{app:rate-osc} we discuss how the amplitude for the neutrino oscillation process derived in the QFT formalism can be related to differential event rates in a neutrino oscillation experiment.

\section{Single-vertex calculation}
\label{sec:single-vertex}

In the QFT approach to neutrino oscillations the combined process of
neutrino production, propagation, and detection is evaluated.
There, we face the unusual situation that two vertices of the process (production and detection) are macroscopically separated in space-time. Before we calculate the amplitude for the oscillation process in \cref{sec:oscAmpl}, we discuss first the conceptually simpler case, where there is only one interaction region, as in standard particle physics interaction calculations. This will serve to introduce the formalism, fix the notation, and to discuss the modifications we introduce to model the specific physical situations of decaying particles as well as particles confined by frequent interactions with the environment.
We consider wave packets for both, initial and final state particles. In \cref{app:rates} we show that, in the limit of plane waves and the appropriate normalization, we can recover standard text book expressions for scattering cross sections and decay rates from our approach.

We define states as superpositions of momentum eigenstates $|\mathbf{k}\rangle$ as
\begin{equation}\label{eq:WP}
  |\phi\rangle = \int d\tilde{\mathbf{k}} \phi(\mathbf{k})|\mathbf{k}\rangle \,,\qquad
  d\tilde{\mathbf{k}} \equiv \frac{1}{2E_k}\frac{d^3k}{(2\pi)^3} \,.
\end{equation}
Three-vectors are denoted by bold-face letters.
Momentum states are normalized as
\begin{equation}
  \langle \mathbf{k'}|\mathbf{k}\rangle = 2E_k(2\pi)^3\delta^{(3)}(\mathbf{k}-\mathbf{k}') 
\end{equation}
and the wave packets as
\begin{equation}
  \int \frac{d^3k}{(2\pi)^3}|\phi(\mathbf{k})|^2 = 1 \,.
\end{equation}

In the following we will consider the specific case of Gaussian wave packets. While this is not true in general, it serves as a useful approximation for the purpose of describing the relevant physics and for order-of-magnitude estimates. The technical advantage is that many integrals can be performed analytically in the case of Gaussian wave packtes. Taking into account the above normalization condition, we have
\begin{equation}
  \phi(\mathbf{k}) = \left(\frac{2\pi}{\sigma^2}\right)^{3/4}e^{-\frac{(\mathbf{k}-\mathbf{p})^2}{4\sigma^2}}\,.
\end{equation}
Here, $\mathbf{p}$ is the mean momentum and $\sigma$ is the momentum spread. We will use the symbol ``$\sigma$'' to denote uncertainties of dimension 1 (i.e., momentum or energy) and the symbol ``$\delta$'' to denote uncertainties of dimension $-1$ (i.e., space or time); they are related via $\sigma\delta = 1/2$.

Let us now consider a process with a set of $N_i$ initial and $N_f$ final state particles, $|\phi_i\rangle$
and $|\phi_f\rangle$, respectively. They are the product of states, each having the form as defined in \cref{eq:WP}. Then, the total transition probability for the process $i\to f$ is given by the $S$-matrix \cite{peskin, weinbergQFT}:
\begin{equation}
  P_{if} = |\langle\phi_f|iT|\phi_i\rangle|^2 \equiv |i\mathcal{A}|^2\,,
\end{equation}
where $iT$ is the non-trivial part of the $S$ matrix: $S=1+iT$. 
We use \cref{eq:WP} and the standard definition of the matrix element for momentum states \cite{peskin},
\begin{equation}
  \left(\prod_f \langle\mathbf{k}_f|\right) iT \left(\prod_i|\mathbf{k}_i\rangle\right) =
  (2\pi)^4\delta^{(4)}\left(\sum_f k_f - \sum_i k_i\right) i\mathcal{M} \,,
\end{equation}
where in general $\mathcal{M}$ depends on all momenta. Then, we obtain the transition amplitude as
\begin{align}
  i\mathcal{A} &=
    \prod_f\int d\tilde{\mathbf{k}}_f \phi_f^*(\mathbf{k}_f)
    \prod_i\int d\tilde{\mathbf{k}}_i \phi_i(\mathbf{k}_i)
    (2\pi)^4\delta^{(4)}\left(\sum_f k_f - \sum_i k_i\right) i\mathcal{M} \label{eq:A0}\\
    &= \int d^4x
    \prod_f\int d\tilde{\mathbf{k}}_f \phi_f^*(\mathbf{k}_f) e^{ik_fx}
    \prod_i\int d\tilde{\mathbf{k}}_i \phi_i(\mathbf{k}_i) e^{-ik_ix} \, i\mathcal{M}  \,.\label{eq:A1}   
\end{align}
In the second step we used the Fourier-transform of the $\delta$-function to introduce an integral over configuration space $d^4x$. This integral will receive contributions from the ``interaction region'' in space-time, determined by the overlap of the wave packets. 

We now proceed by performing the momentum integrals by adopting the following approximations.
First, we assume that the matrix element $\mathcal{M}$ varies slowly with momenta on the scales $\sigma_{i,f}$ and we can replace $\mathcal{M}(\mathbf{k}_i, \mathbf{k}_f)\approx \mathcal{M}(\mathbf{p}_i, \mathbf{p}_f)$. Then it can be pulled out of the integrals.
Second, consider the exponentials $e^{ik_{i,f}x}$. For a generic $k_{i,f}$ we have
$kx = E_k t - \mathbf{kx}$. Again we assume that the momentum is not too far from its mean value $\mathbf{p}$ and approximate 
\begin{align}\label{eq:expandE}
  E_k \approx E_p + \mathbf{v}(\mathbf{k} - \mathbf{p}) 
\end{align}
with the group velocity
\begin{align}\label{eq:v}
  \mathbf{v} \equiv \left. \frac{\partial E_k}{\partial \mathbf{k}} \right|_{\mathbf{k}=\mathbf{p}} \,.
\end{align}
The physical implication of the approximation \cref{eq:expandE} is that the wave packets are not spreading in time.\footnote{Wave packet dispersion would be described by the second order term in the expansion \cref{eq:expandE}, which is of order $\sigma^2/E_p$. This term is small compared to typical energy widths $\sigma_m, \sigma_E,\Gamma_{\rm col}$ derived below, because $\sigma/E_p \ll 1$. Physically this means that on time scales relevant for the coherent interaction the spreading of wave packets can be neglected.}
Stationary states correspond to $\mathbf{v}=0$, whereas for free particles we have $\mathbf{v} = \mathbf{p}/E_p$. At this point we do not need to specify the dispersion relation and can allow for arbitrary $\mathbf{k}$ dependence of $E_k$.

With these two approximations all integrals become integrals over Gaussians, which can be performed, see e.g., \cite{Beuthe:2001rc}.
The momentum integrals are all of the type
\begin{align}\label{eq:N}
  \int d\tilde{\mathbf{k}} \phi(\mathbf{k}) e^{\mp ikx} \approx
  \mathcal{N} e^{\mp i px - (\mathbf{x}-t\mathbf{v})^2 \sigma^2} \,,\qquad
  \mathcal{N} \equiv \frac{1}{\sqrt{2E_{p}}} \frac{(2\sigma)^{3/2}}{(2\pi)^{3/4}} \,.
\end{align}
Using this in \cref{eq:A1}, we obtain
\begin{align}
  i\mathcal{A} = i\mathcal{M} \int d^4x \prod_{i,f} \mathcal{N}_i \mathcal{N}_f
  \exp\left[i\left(\sum_f p_f - \sum_i p_i\right)x -
    \sum_{i,f}(\mathbf{x}-t\mathbf{v}_{i,f})^2 \sigma^2_{i,f}\right] \,.
\end{align}
In order to simplify the exponential we introduce some notation:
\begin{align}
  \Delta E \equiv \sum_i E_{p_i} - \sum_f E_{p_f}  \,,\qquad
  \Delta \mathbf{p} \equiv \sum_i \mathbf{p}_i -\sum_f \mathbf{p}_f \,, 
\end{align}
such that $\Delta E = 0$ and $\Delta\mathbf{p}=0$ correspond to exact energy and momentum conservation, respectively.
We define the total momentum spread by
\begin{align}\label{eq:sigmap}
  \sigma_p^2 \equiv \sum_{i,f}\sigma^2_{i,f} \,,
\end{align}
and a weighted velocity and velocity-squared:
\begin{align}\label{eq:v_eff}
  \mathbf{v} \equiv \frac{1}{\sigma_p^2} \sum_{i,f} \sigma_{i,f}^2 \mathbf{v}_{i,f} \,,\qquad
  \Sigma \equiv \frac{1}{\sigma_p^2} \sum_{i,f} \sigma_{i,f}^2 \mathbf{v}_{i,f}^2 \,.
\end{align}
Using these definitions to rewrite the argument of the exponential,
the $d^4x$ integration can be performed as well:
\begin{align}
  i\mathcal{A} &= i\mathcal{M} \prod_{i,f} \mathcal{N}_i \mathcal{N}_f
  \int d^4x \,
  \exp\left[-i(\Delta E t - \Delta\mathbf{px})  - \sigma_p^2(\mathbf{x}^2 - 2\mathbf{xv}t + \Sigma t^2)\right]
  \label{eq:d4x}\\
  &=
  i\mathcal{M} \prod_{i,f} \mathcal{N}_i \mathcal{N}_f \,
   \frac{\pi^2}{\sigma_p^3\sigma_e}
  \exp\left[ -\frac{(\Delta\mathbf{p})^2}{4\sigma_p^2}
    -\frac{(\Delta E - \Delta\mathbf{pv})^2}{4\sigma_e^2}\right] \,, \label{eq:A2}
\end{align}
where the effective energy spread is obtained as
\begin{align}\label{eq:sigmae}
  \sigma_e^2 \equiv \sigma_p^2(\Sigma - \mathbf{v}^2) \,.
\end{align}
We notice that the quantities $\sigma_e$ and $\sigma_p$ correspond to the amounts within which energy and momentum conservation can be violated. It is easy to see that they fulfill
%
$0 \le \sigma_e \le \sigma_p$.
%
The momentum uncertainty, \cref{eq:sigmap}, is just the sum of the squares of the
momentum uncertainty of all particles. Hence, it is dominated by the
particle with the largest uncertainty. For hierarchical uncertainties, $\sigma_e$ is generally dominated by the particle with the second largest uncertainty, see e.g., \cite{Beuthe:2001rc} for a discussion. Note that ``energy-momentum violation'' happens for the \textit{mean} quantities $p_i,p_f$, whereas for the actual 4-momenta picked at the interaction vertex, energy-momentum conservation is exact, as manifest by the $\delta$-function in \cref{eq:A0}. Therefore, it is not surprising that the mean quantities need to fulfill 4-momentum conservation only approximately.

\subsection{Interactions with the environment and particle decay}
\label{sec:interact}

We consider now the case that the involved particles propagate through
a medium, with frequent collisions. When an interaction happens, the particle wave function picks up a random phase, which leads to loss of coherence. We model this effect by restricting the time integration in \cref{eq:d4x} from $t=0$ till the time when the next interaction happens by introducing an exponential $e^{-t\Gamma_{\rm col}/2}$, where $\Gamma_{\rm col}$ is the collision rate. This is motivated by considering the scattering as a Poissonian process.\footnote{See also \cite{grimus} for a similar approach.} Hence, the integral in \cref{eq:d4x} becomes
\begin{align}
  \int d^4x \, e^{-i(\Delta E t - \Delta\mathbf{px})  - \sigma_p^2(\mathbf{x}^2 - 2\mathbf{xv}t + \Sigma t^2)}
  e^{-t\Gamma_{\rm col}/2}\Theta(t)
  &= 
   \frac{\pi^{3/2}}{\sigma_p^3}
   e^{-\frac{(\Delta\mathbf{p})^2}{4\sigma_p^2}}
   \int_0^\infty dt \, e^{-\sigma_e^2t^2 - it\Omega - t\Gamma_{\rm col}/2} \label{eq:d4x2}\\
   &\to \frac{\pi^{3/2}}{\sigma_p^3}
   e^{-\frac{(\Delta\mathbf{p})^2}{4\sigma_p^2}}
   \frac{1}{i\Omega + \Gamma_{\rm col}/2}\,, \label{eq:lorentzian}
\end{align}
with $\Omega\equiv \Delta E - \Delta\mathbf{pv}$, and in the last step
we assumed $\sigma_e \ll \Gamma_{\rm col}$. Hence, after squaring the amplitude, instead of the energy Gaussian factor in \cref{eq:A2} we obtain the typical Lorentzian shape $\propto 1/(\Omega^2 + \Gamma_{\rm col}^2/4)$. This modelling describes correctly the collisional line broadening \cite{VanVleck:1945zz}, see \cref{app:decay}.

Actually, we will take this picture as the physical origin of wave
packets for our external particles: the interactions with the
environment lead to the localization of the particle within a length
$\delta_x$ which in turn determines the momentum spread $\sigma$ via
$\sigma \delta_x = 1/2$ as well as the mean time between two scatterings as
$\delta_x / |\mathbf{v}|$. We adopt
now the following ansatz for the effective collision rate $\Gamma_{\rm col}$
\begin{equation}\label{eq:Gamma}
  \Gamma_{\rm col}^2 \equiv \sum_{i,j} \sigma_{i,f}^2 \mathbf{v}_{i,f}^2 = \sigma_p^2 \Sigma \,,
\end{equation}
i.e., adding the interaction rates in squares.\footnote{The reason for this ansatz is our Gaussian approximation, see below, where naturally all spreads are added quadratically. An alternative definition could be to add them linearly, $\tilde{\Gamma}_{\rm col} = \sum_{i,f} \sigma_{i,f} |\mathbf{v}_{i,f}|$. It is easy to show that $\tilde{\Gamma}_{\rm col} \ge \Gamma_{\rm col}$. Numerically we find with the numbers from \cref{sec:numerics} that in our cases of interest we have $\tilde{\Gamma}_{\rm col} \approx \Gamma_{\rm col}$.}
Hence, $\Gamma_{\rm col}$ is dominated by the particle with the fastest
interaction rate: as soon as one of the involved particles interacts
with the environment, coherence is lost. Comparing \cref{eq:Gamma}
with the expression for the energy spread in \cref{eq:sigmae}, we see
that $\sigma_e \le \Gamma_{\rm col}$. In \cref{sec:numerics} we will find for our cases of interest that $\sigma_e \approx \Gamma_{\rm col}$ because in most cases $\mathbf{v}^2 \ll \Sigma$. 

As we have seen, our physical picture leads to a constraint on $\Omega$ with Lorentzian shape. In the following we will, however, replace the Lorenzian with a corresponding Gaussian factor with width $\Gamma_{\rm col}$. This approximation will still capture the relevant physics, but simplify the calculations significantly, because integrals can be taken analytically. Hence, we recover an expression for the amplitude, similar to eq. \eqref{eq:A2}, but with the replacement
\begin{equation}
  \sigma_e^2 \to  \sigma_e^2 + \Gamma_{\rm col}^2 = \sigma_p^2(2\Sigma - \mathbf{v}^2) \,.
\end{equation}

In neutrino oscillation experiments neutrinos are very often produced in particle decays, e.g., beta-decay of a nucleus in reactor experiments or pion decay in accelerator experiments. We can model the decay of a particle in complete analogy to the interaction with the environment by introducing an additional factor $e^{-t\Gamma_{\rm dec}/2}$ in 
\eqref{eq:d4x2}. Hence, the width of the Lorentzian becomes just $\Gamma_{\rm col}+\Gamma_{\rm dec}$. As we show in \cref{app:decay}, in the limit $\Gamma_{\rm col} \gg \Gamma_{\rm dec}$ (frequent collisions) this ansatz describes the collisional line broadening of a decay line, whereas in the limit $\Gamma_{\rm col} \ll \Gamma_{\rm dec}$ (fast decay and negligible collisions) it reproduces the standard definition of the decay width in terms of the matrix element. In the Gaussian approximation we simply obtain the effective energy spread by
\begin{equation}\label{eq:sigmaE}
  \sigma_e^2 \to \sigma_E^2 \equiv \sigma_e^2 + \Gamma_{\rm col}^2 + \Gamma_{\rm dec}^2 = \sigma_p^2(2\Sigma - \mathbf{v}^2)  + \Gamma_{\rm dec}^2 \,.
\end{equation}
For neutrino production in a nuclear reactor as well as the source in Gallium experiments, the lifetimes of the relevant beta decays are typically much longer than the interaction times with the environment (see \cref{sec:numerics}), and therefore, it is save to negect $\Gamma_{\rm dec}$ compared to $\Gamma_{\rm col}$. This may not be the case for experiments using pion decay, where pions decay in a decay tunnel essentially without interacting.

Although we can combine the energy spread $\sigma_e$, the collision rate, and the decay rate in the same effective quantity $\sigma_E$ (thanks to the Gaussian approximation for all of them) their physical origins are different. $1/\sigma_e$ corresponds to the effective interaction time due to the wave packet overlap of all involved particels, $\Gamma_{\rm col}$ describes the coherence loss due to interactions of each of the involved particles with the environment, and $\Gamma_{\rm dec}$ takes into account the energy spread due to the finite lifetime of a decaying particle.

\section{Neutrino oscillation amplitude and event rate}
\label{sec:oscAmpl}

\begin{figure}[t]
	\centering
	\begin{tikzpicture}
	\begin{feynman}
	\def\a{2.5cm}
	\def\m{2.5*0.707cm}
	\vertex [blob] (l) at (0,0){};
	\vertex [blob] (r) at (3*\a,0){};
	\vertex at (-\a,0) (a){$p_A$};
	\vertex at (3*\a-\m,-\m) (b){$p_B$};
	\vertex at (\m,\m) (f1){$p_1$};
	\vertex at (\m,-\m)(f2){$p_2$};
	\vertex at (3*\a+\m,\m) (f3){$p_3$};
	\vertex at (3*\a+\m,-\m) (f4){$p_4$};
	\diagram* {
		(l)--[anti fermion, edge label=$\bar\nu$](r),
		(l)--[anti fermion](a),
		(r)--[anti fermion](b),
		(l)--[fermion](f1),
		(l)--[fermion](f2),
		(r)--[fermion](f3),
		(r)--[anti fermion](f4),};
	\end{feynman}
	\end{tikzpicture}
	\mycaption{Feynman diagram for the total process in an oscillation experiment.}
	\label{fig:diagram}
\end{figure}

We now move to the discussion of the amplitude relevant for neutrino oscillation experiments consisting of neutrino production, propagation and detection. To be
specific, we consider neutrino production by the decay of a particle
$A$ into two final state particles and an anti-neutrino, $A\to
1+2+\bar\nu$, and anti-neutrino detection via the process $B + \bar\nu \to 3+4$.
We have in mind reactor neutrinos, where the production process
corresponds to the beta decay of a nucleus ($A$), and the detection
process is the inverse beta decay reaction on a proton ($B$), but many
of our considerations will apply also in other circumstances with
minor modifications. The total process $A+B \to 1+2+3+4$ is
illustrated in \cref{fig:diagram}. The neutrino is considered as an
internal propagtor and does not appear as external particle \cite{Rich:1993wu,Giunti:1993se,Grimus:1996av}.  We
now follow the common approach \cite{Beuthe:2001rc} and calculate the
amplitude for the total process, by assuming wave packets for all
external particles, both initial state ($A,B$) as well as final state
particles ($1,2,3,4$).

We proceed in complete analogy to the discussion in
\cref{sec:single-vertex} but generalize it to the case of
macroscopically separated production ($P$) and detection ($D$) regions.
In analogy to \cref{eq:A2}, we obtain the following expression for the amplitude describing production of an anti-neutrino with flavour $\alpha$ and detection of an anti-neutrino with flavour $\beta$\footnote{In the case of reactor neutrino experiments we have of course $\alpha=\beta=e$.} (see e.g., 
\cite{Beuthe:2001rc,Akhmedov:2010ms} for explicit derivations):
\begin{align}
  i\mathcal{A}_{\alpha\beta} =& \sum_j U_{\alpha j}U_{\beta j}^* \left(\prod_{i=A,B,f} \mathcal{N}_i\right)
  \int \frac{d^4p}{(2\pi)^4}
  i\tilde{\mathcal{M}}_P \frac{\slashed{p} - m_j}{p^2-m_j^2 + i\epsilon} i\tilde{\mathcal{M}}_D \,e^{-ip(x_D-x_P)}
  \nonumber\\
  &\times \prod_{I=P,D} 
   \frac{\pi^2}{\sigma_{pI}^3\sigma_{EI}}
  \exp\left[ -\frac{(\mathbf{p} - \mathbf{p}_I)^2}{4\sigma_{pI}^2}
    -\frac{(p^0 - E_{I} - \mathbf{v}_I (\mathbf{p} - \mathbf{p}_I))^2}{4\sigma_{EI}^2}\right] \,. \label{eq:A3}
\end{align}
Here, $U_{\alpha j}$ are elements of the PMNS mixing matrix, the normalization factors $\mathcal{N}_i$ are defined in \cref{eq:N}, $\tilde{\mathcal{M}}_{P,D}$ are the reduced matrix elements of the production and detection processes, the sum over $j$ runs over the neutrino mass states with neutrino mass $m_j$, $x_{P,D}$ are space-time points located in the production and detection region. In the second line of \cref{eq:A3} we obtain two Gaussian factors related to the approximate energy-momentum conservation at production and detection points, with the momentum spreads $\sigma_{pI}$, energy spreads $\sigma_{EI}$, and velocities $\mathbf{v}_I$ (with $I=P,D$) defined as in \cref{eq:sigmap}, \cref{eq:sigmaE}, \cref{eq:v_eff}, respectively. Furthermore, we have defined the kinematic 4-momenta of the neutrino at the production and detection vertices:
\begin{align}
  p_P &= p_A - p_1 - p_2 \,,  \label{eq:EP}\\
  p_D &= -p_B + p_3 + p_4 \,, \label{eq:ED}
\end{align}
and $E_{P},E_D$ are the time-components of the corresponding 4-vectors.

Next we perform the integral over $d^3p$ by
using the Grimus-Stockinger theorem \cite{Grimus:1996av}, which allows
us to take into account the macroscopic separation of source and
detector. In the relevant limit the propagating neutrinos go on-shell, and we obtain
\begin{align}
  i\mathcal{A}_{\alpha\beta} =& \sum_j \frac{U_{\alpha j}U_{\beta j}^*}{4\pi L}
  \left(\prod_{i=A,B,f} \mathcal{N}_i\right)
  i\mathcal{M}_P  i\mathcal{M}_D \,
   \frac{\pi^4}{\sigma_{pP}^3\sigma_{EP}\sigma_{pD}^3\sigma_{ED}}
  \nonumber\\
  &\times
  \int \frac{dp^0}{2\pi}
  \exp\left[ -ip^0T + ip_jL -f_j(p^0) \right] \,, \label{eq:A4}
\end{align}
where $\mathbf{L} = \mathbf{x}_D - \mathbf{x}_P$, $L = |\mathbf{L}|$,
$\hat{\mathbf{l}} = \mathbf{L}/L$, $T = t_D - t_P$.
The function $f_j(p^0)$ is obtained from the exponential in the second line of \cref{eq:A3}, where according to the Grimus-Stockinger theorem $\mathbf{p} \to \mathbf{p}_j$ with
\begin{align}
  \mathbf{p}_j = p_j \hat{\mathbf{l}} \,,\quad p_j = \sqrt{(p^0)^2 - m_j^2} \,.
\end{align}
Following Ref.~\cite{Beuthe:2001rc},
we rewrite $f_j(p^0)$ by decomposing the vectors in components parallel and orthogonal to $\hat{\mathbf{l}}$:
\begin{align}\label{eq:f}
  f_j(p^0) = \sum_{I=P,D} \left[
  \frac{(p_j - p_I)^2}{4\sigma_{pI}^2}
    + \frac{(p^0 - E_{I} - v_I (p_j - p_I))^2}{4\sigma_{EI}^2}
   + \frac{\mathbf{p}_{I\perp}^2 }{4\sigma_{pI}^2} \right] \,,
\end{align}
where $p_I$ and $v_I$ denote the components parallel to $\hat{\mathbf{l}}$ and we have redefined
\begin{align}\label{eq:Enew}
   E_I - \mathbf{v}_{I\perp} \mathbf{p}_{I\perp} \to E_I \,.
\end{align}
In order to keep notation concise we do not introduce a different symbol for this new variable.

\subsection{Derivation of the decoherence terms}
\label{sec:decoh}

\Cref{eq:A4} is the starting point to derive the terms leading to decoherence effects, which will be the main focus of our considerations in the following. We are not concerned with overall factors and focus on the interference terms. First, we have to square the amplitude and perform an integration of the unobservable propagation time $T$, see discussion in \cref{app:rate-osc}, \cref{eq:Asq-average}. We obtain a $\delta$-function from the integral $\int dT e^{-i(p^0-{p'}^0)T}$ and obtain
\begin{align} \label{eq:Asq1}
  \overline{|\mathcal{A}_{\alpha\beta}|^2} \propto \int dT |\mathcal{A}_{\alpha\beta}(T)|^2 \propto
  \sum_{jk} U_{\alpha j}U_{\beta j}^*U_{\alpha k}^*U_{\beta k}
  \int dp^0 \exp\left[ i\frac{\Delta m^2_{kj}L}{2p^0} - f_j(p^0)-f_k(p^0) \right]\,,
\end{align}
where we used that neutrino masses are small, $m_j \ll p^0$, and
expand the square root in the neutrino momenta as $p_j \approx p^0 - m_j^2/(2p^0)$.
We see that the time integration implies that
only neutrinos with the same energy can interfere and different
neutrino energies are summed incoherently
\cite{Beuthe:2001rc,Akhmedov:2010ms}.

The interference terms correspond to $j \neq k$ in \cref{eq:Asq1}, for
which we recognize the familiar oscillation phase depending on $\Delta
m^2_{kj} \equiv m_k^2- m_j^2$. Let us now simplify the discussion and
specialize to the case of two neutrino states with masses $m_1, m_2$,
and write
\begin{align}
  m_1^2 = \overline{m}^2 - \frac{1}{2} \Delta m^2 \,,\quad
  m_2^2 = \overline{m}^2 + \frac{1}{2} \Delta m^2 \,.
\end{align}
We are interested only in terms proportional to $\Delta m^2$. Terms depending only on the absolute neutrino mass $\overline{m}^2$ will lead to (tiny) irrelevant global corrections which do not affect the interference term. Furthermore, the quantities $m_I^2 = E_I^2 - p_I^2$ (for $I=P,D$) are of order of the neutrino mass-squared $\overline{m}^2$ and also independent of the neutrino mass indices $j,k$. Therefore, we neglect also terms proportional to $m_I^2$ and set $E_I \approx p_I$. Using $p_j \approx p^0 - m_j^2/(2p^0)$  also in the function $f_j(p^0)$ and dropping all terms proportional to $\overline{m}^2$ and independent of $\Delta m^2$, we obtain at leading order in $\Delta m^2$:
\begin{align}
  f_1(p^0)+f_2(p^0) = \sum_{I=P,D} \frac{(p^0 - E_I)^2}{2 \sigma_{I,\rm eff}^2}
  + \frac{1}{2}\left(\frac{\Delta m^2}{4p^0\sigma_m}\right)^2
  + \sum_{I=P,D} \frac{\mathbf{p}_{I\perp}^2 }{2\sigma_{pI}^2} \label{eq:f12}
\end{align}
with
\begin{align}
  \frac{1}{\sigma_{I,\rm eff}^2} &\equiv  \frac{1}{\sigma_{pI}^2} + \frac{(1 - v_I)^2}{\sigma_{EI}^2} \,,  \label{eq:sigma_eff}\\
  \frac{1}{\sigma_m^2}         &\equiv  \sum_{I=P,D} \left(\frac{1}{\sigma_{pI}^2} + \frac{v_I^2}{\sigma_{EI}^2} \right)\,. \label{eq:sigmam}
\end{align}
Already at this stage we obtain the term with $\sigma_m$ which potentially can lead to decoherence.
We will comment on its physical interpretation in \cref{sec:decoh-discussion}.

Next we perform the integral over $p^0$. We expand the first term in \cref{eq:f12} around its minimum, which is at $p^0 = E_0$ with
\begin{align}
  E_0 \equiv  \sigma_{\rm eff}^2 \sum_{I=P,D} \frac{E_I}{\sigma_{I,\rm eff}^2} \,,\quad 
  \frac{1}{\sigma_{\rm eff}^2} \equiv \sum_{I=P,D} \frac{1}{\sigma_{I,\rm eff}^2} \,. \label{eq:E0}
\end{align}
The oscillatory phase $i\Delta m^2L/(2p^0)$ is expanded to first order
in $(p^0-E_0)$, and we set $p^0=E_0$ in the term with $\sigma_m$
(ignoring higher order correction to this decoherence term).
Then, the $p^0$-integral can be performed with the method outlined in \cref{app:int} and we obtain
\begin{align}
  \overline{|\mathcal{A}_{\alpha\beta}|^2} \propto&
  \exp\left[i\frac{\Delta m^2L}{2E_0}\right] \times
  \exp\left[ - \sum_{I=P,D} \frac{\mathbf{p}_{I\perp}^2 }{2\sigma_{pI}^2} \right]  
  \nonumber\\ 
  &\times \exp\left[
    -\frac{1}{2} \left(\frac{\Delta m^2}{4E_0\sigma_m}\right)^2
    -\frac{1}{2} \left(\frac{\Delta m^2 L \sigma_{\rm eff}}{2E_0^2}\right)^2
    -\frac{1}{2} \frac{(E_D-E_P)^2}{\sigma_{P,\rm eff}^2+\sigma_{D,\rm eff}^2}
    \right]  \label{eq:Asq2}
\end{align}
This is a central result of the QFT approach to the oscillation process; let us briefly comment on the terms appearing here. In the first line we have the standard oscilation phase and a term depending on the net momentum components orthogonal to the neutrino direction. The latter term will constrain the phase space integrals for the orthogonal components within the mometum spreads $\sigma_{pI}$ around zero. Because of the relabeling of the energy variables $E_{P,D}$ in \cref{eq:Enew}, also the oscillation phase depends on $\mathbf{p}_{I\perp}$ and in principle the integral over these components can lead to additional non-trivial effects (we comment on it in \cref{app:phase-space}).

Note that in \cref{eq:Asq2} we have three quantities corresponding to an effective neutrino energy: the kinematic neutrino energies $E_{P}$ and $E_{D}$ at the production and detection vertices defined in \cref{eq:EP,eq:ED}, respectively, and $E_0$ which is a weighted mean value of the former two. The last term in \cref{eq:Asq2} ensures, that all ``three neutrino energies'' are the same within quantum mechanical uncertainties, determined by the sum of the effective energy-momentum uncertainties at source and detector. \Cref{eq:Asq2} is completely symmetric with respect to source and detector.

\subsection{Discussion of the decoherence terms}
\label{sec:decoh-discussion}

The first two terms in the second line of \cref{eq:Asq2} describe the exponential damping due to decoherence.  We provide here a brief review of the two decoherence terms, see e.g., Refs.~\cite{Beuthe:2001rc,Akhmedov:2009rb,giunti,Akhmedov:2010ms} for more discussions. The two terms correspond to two generic types of damping. We define
\begin{align}
  \xi_{\rm loc} & = \exp\left[
    -\frac{1}{2} \left(\frac{\Delta m^2}{4E_\nu\sigma_m}\right)^2 \right]
  = \exp\left[
    -\frac{1}{2} \left(\frac{\Delta m^2\delta_{\rm loc}}{2E_\nu}\right)^2 \right] \,,
  \label{eq:xi_loc} \\
  \xi_{\rm en} &=
  \exp\left[
    -\frac{1}{2} \left(\frac{\Delta m^2 L}{2E_\nu} \frac{\sigma_{\rm en}}{E_\nu}\right)^2 \right]
  = \exp\left[
    -\frac{1}{2} \left(\frac{\Delta m^2 L}{4E_\nu^2\delta_{\rm en}} \right)^2 \right] \,,
\label{eq:xi_en}
\end{align}
with
\begin{align}
  \sigma_m\delta_{\rm loc} = \frac{1}{2} \,,\qquad
  \sigma_{\rm en}\delta_{\rm en} = \frac{1}{2} \,.
\end{align}
We introduced the generic energy spread $\sigma_{\rm en}$, where in \cref{eq:Asq2} we have 
$\sigma_{\rm en} = \sigma_{\rm eff}$, and $E_\nu$ is a relevant neutrino energy, $E_0$ in the version obtained in \cref{eq:Asq2}.

Both decoherence terms have a simple physical interpretation. 
Starting with $\xi_{\rm loc}$, the interpretation in the energy-representation of this term is that the energy-momentum uncertainty encoded in $\sigma_m$ needs to be large enough, such that individual mass states can neither be resolved at the production nor at the detection process. If energy-momentum was defined with an accuracy better than $\Delta m^2/E_\nu$, the individual neutrino mass states would be determined and no interference of different mass states would be possible.

Another interpretation of this term becomes apparent in the spatial representation: defining the oscillation length by
\begin{align}
  L_{\rm osc} = 2\pi \frac{2E_\nu}{\Delta m^2} 
\end{align}
the decoherence terms can be written as
\begin{align} 
  \xi_{\rm loc} &= \exp\left[-2\pi^2\left(\frac{\delta_{\rm loc}}{L_{\rm osc}}\right)^2\right] \,,
  \label{eq:xi_loc2}\\
  \xi_{\rm en} &= \exp\left[-2\pi^2\left(\frac{L}{L_{\rm osc}}\frac{\sigma_{\rm en}}{E_\nu}\right)^2\right]
  \label{eq:xi_en2} \,.
\end{align}
In this form we see that $\xi_{\rm loc}$ can also be interpreted as the condition $\delta_{\rm loc}\ll L_{\rm osc}$, i.e., that both, source and detection points need to be localized better than the oscillation length, by noting that $\delta_{\rm loc}^2 = \delta_P^2+\delta_D^2$.

Moving now to $\xi_{\rm en}$, this term says that for an experiment around the oscillation maximum ($L/L_{\rm osc} \simeq 1$), the neutrino energy needs to be sufficiently determined, such that $\sigma_{\rm en} \ll E_\nu$; for experiments beyond the oscillation maximum the condition becomes correspondingly stronger. Using $\sigma_{\rm en} = \sigma_{\rm eff}$ with $\sigma_{\rm eff}$ defined in \cref{eq:sigma_eff,eq:E0}, this is a condition on the energy and momentum spreads of external particles of the neutrino production and detection processes. For $L\simeq L_{\rm osc}$, decoherence will become relevant only if $\sigma_{\rm en} \simeq E_\nu$, i.e., the quantum mechanical uncertainty on the neutrino energy needs to become comparable to the neutrino energy itself, implying that the neutrino would not have a well defined energy. Note that we work under the assumption that momentum spreads are small compared to relevant momenta or energies, and therefore our approximations adopted in \cref{sec:single-vertex} to perform the integrals may not apply if $\sigma_{\rm en} \sim E_\nu$.

In the above derivation we have integrated first over $T$ and then over $p^0$, which is more convenient to derive decoherence terms. If the order of the $T$ and $p^0$ integrals are exchanged, one can see that an exponential factor makes sure that the $T$ integral is dominated by values of $T$ constrained by $|T-L-Lm_j^2/(2E_0^2)| \lesssim 1/\sigma_{\rm eff}$, which can be interpreted as relating $L \simeq T v_j$, with $v_j\approx 1 - m_j^2/(2 E_0^2)$ corresponding to the ``velocity'' of the neutrino with mass $m_j$ \cite{Beuthe:2001rc,Raphael-MSc}. This suggests a wave packet interpretation of the internal neutrino in the QFT approach \cite{Akhmedov:2010ms}. In that picture the damping due to $\xi_{\rm en}$ can be interpreted as wave packet separation of the propagating neutrinos.

We note that the two terms $\xi_{\rm loc}$ and $\xi_{\rm en}$ have the opposite dependence on the spreads $\sigma_m$ and $\sigma_{\rm en}$. This means that for oscillations to be observable, quantum uncertainties have to be big enough that different mass states can interfere ($\sigma_m \gg \Delta m^2/E_\nu$), but small enough that interference is not damped ($\sigma_{\rm en} \ll E_\nu L_{\rm osc}/L$). Assuming that very roughly $\sigma_m \sim \sigma_{\rm en}$, we see that there are many orders of magnitude available to fulfill both requirements, thanks to the smallness of $\Delta m^2/E_\nu^2$ or, in other words, due to the macroscopically large oscillation length, $L_{\rm osc} E_\nu \gg 1$.

\subsection{Phase-space integrals over unobserved external momenta}
\label{sec:phase-space}

\Cref{eq:Asq2} contains the intrinsic quantum mechanical decoherence terms. So far we have only averaged the amplitude squared over the unobservable time $T$ (which actually does not introduce a decoherence term). All other manipulations are performed at the amplitude level; in particular, the integral over $p^0$ corresponds to an internal particle, which can be performed already at amplitude level. In real experiments there are of course always effects leading to additional averaging at the probability level, i.e., of the amplitude squared (see also \cref{sec:classical}). Some of these averages are intrinsically unavoidable and related to the physical configuration of the experiment.

Let us consider first the case of reactor neutrino oscillation experiments; a very similar discussion applies also to accelerator or atmospheric neutrino experiments. In
these experiments usually the neutrino energy is reconstructed in the
detector, by measuring all (or some of) the outgoing particles at the
detector, i.e., particles 3 and 4 in our example, which allows to
reconstruct $E_D$ via \cref{eq:ED,eq:Enew} with some accuracy. In
contrast, inital and final state particles at the production point
usually are not observed.\footnote{We do not consider here so-called
  monitored neutrino beams~\cite{Longhin:2022tkk}, which (at least in
  principle) would allow also to reconstruct $E_P$.}  Therefore, in
order to calculate event rates $R_D$ in the detector,
we need to integrate the squared amplitude over the phase
space of final state particles in the production reaction, see also
\cref{app:rate-osc}. By a suitable variable transformation, one of
these integrals can be chosen to be over $E_P$. As we sketch in \cref{app:phase-space}, the decoherence term emerging from this integral has the same shape as $\xi_{\rm en}$ and can be combined with the original term present already in \cref{eq:Asq2} such that we obtain for the event rate in the detector
\begin{align}
  R_D(L,E_D) \propto \int dE_P \overline{|\mathcal{A}_{\alpha\beta}|^2} & \propto
   \exp\left[i\frac{\Delta m^2L}{2E_D}
    -\frac{1}{2} \left(\frac{\Delta m^2}{4E_D\sigma_m}\right)^2
    -\frac{1}{2} \left(\frac{\Delta m^2 L \sigma_{D,\rm eff}}{2E_D^2}\right)^2
    \right] \,.
 \label{eq:R_D}  
\end{align}
Hence, the energy spread in $\xi_{\rm en}$ is given by $\sigma_{\rm en} = \sigma_{D,\rm eff}$, i.e., it depends only on the uncertainties in the detector, while $\sigma_m$ in $\xi_{\rm loc}$ remains unchanged and contains contributions from both, the production and detection process, see \cref{eq:sigmam}.

The integral over $E_P$ corresponds to a \emph{classical} sum, i.e., summing the squared amplitude. The corresponding decoherence can therefore be considered as emerging from classical averaging. Note, however, here this averaging is in principle unavoidable, given the physical observables in a neutrino oscillation experiment. In this way, \cref{eq:R_D} depends only on the ``observable neutrino energy'' $E_D$. Also note that $\sigma_{D,\rm eff} \ge \sigma_{\rm eff}$. Therefore, the classical averaging due to the $E_P$ integration increases effectively the decoherence, see \cref{sec:classical}. The fact that only the detection process determines the decoherence in the last term of \cref{eq:R_D} and all production-related uncertainties drop out follows from expanding the oscillation phase around $E_D$ and from the Gaussianity of all involved uncertainties; this result may not hold in the most general case. Note that we keep only leading terms in the expansion parameters $\sigma/E$ with $\sigma \in (\sigma_{\rm eff},\sigma_{P,\rm eff},\sigma_{D,\rm eff})$ and $E \in (E_0,E_P,E_D)$, and differences like $\sigma_{D,\rm eff}/E_D - \sigma_{P,\rm eff}/E_P$ are of higher order in this expansion. However, our result that the relevant energy spread in the $\xi_{\rm en}$ term of \cref{eq:R_D} is larger than $\sigma_{\rm eff}$ as in \cref{eq:Asq2} is robust.

In typical detectors of modern reactor experiments only the energy of the outgoing positron can be determined, whereas the neutron momentum is not measured. Hence, we need an additional phase-space integral over the neutron momentum. This will provide another contribution to $\sigma_{\rm en}$, which however, is at most of the same order of magnitude and typically much smaller (see the discussion of the $\mathbf{p}_\perp$ integral in \cref{app:phase-space}).

Moving to Gallium radioactive source experiments, these are pure counting experiments, i.e., the detector does not provide energy information, whereas the neutrino source consists of a couple of quasi-monochromatic neutrino lines from an electron-capture decay $N \to N' + \nu$. In this case one would naturally integrate the phase space over $E_D$ instead of $E_P$. The calculation is completely symmetric to the one outlined above and we obtain the same result as in \cref{eq:R_D} with the replacements $E_D\to E_P$ and $\sigma_{D,\rm eff} \to \sigma_{P,\rm eff}$. Hence, in this case the energy spread in $\xi_{\rm en}$ is set by the production process and will be determined by the localization of the nuclei $N,N'$ as well as the (natural and/or thermal) linewidth of the decay (see \cref{sec:numerics}). The same comments as above apply regarding the dropping out of $\sigma_{D,\rm eff}$.

\subsection{Classical averaging} 
\label{sec:classical}

In realistic experiments both production point and detection point are known only within some uncertainty, related to the size of the neutrino source and the vertex resolution of the detector. Similarly, detectors can determine particle energy and momentum only within certain resolutions, and hence the neutrino energy $E_D$ can be reconstructed only within a finite accuracy. These effects are taken into account in predicted event rates by convoluting the neutrino oscillation probability with the corresponding resolution functions. Hence, a classical average is performed.

Let us approximate these spatial and energy resolutions by Gaussians with widths $\delta_{\rm clas}$ and $\sigma_{\rm clas}$, correspondingly: 
\begin{align}
  &\int dL' \, R_D(L',E_\nu) \frac{1}{\sqrt{2\pi}\delta_{\rm clas}} \exp\left[-\frac{(L'-L)^2}{2\delta_{\rm clas}^2}\right] \,,\\
  &\int dE'_D \, R_D(L,E'_D) \frac{1}{\sqrt{2\pi}\sigma_{\rm clas}} \exp\left[-\frac{(E'_D-E_D)^2}{2\sigma_{\rm clas}^2}\right] \,,
\end{align}
with $R_D(L,E_D)$ given in \cref{eq:R_D}. Assuming that the width of the Gaussians is small compared to the other $L$ or $E_D$ dependent factors in $R_D$, we can expand the oscillation phase to linear order either in $(L'-L)$ or $(E'_D - E_D)$ and see that the integral takes again the same form as in \cref{app:int}. Applying \cref{eq:int-app}, we obtain decoherence terms of the same form as $\xi_{\rm loc}$ for the $L$ smearing and $\xi_{\rm en}$ for the energy resolution. Hence, they can be combined with the corresponding terms present in \cref{eq:R_D}, which amounts to the replacement
\begin{align}
  \delta_{\rm loc}^2 &\to  \delta_{\rm loc}^2 + \delta^2_{\rm clas} \,,\qquad
  \sigma_{D,\rm eff}^2 \to  \sigma_{D,\rm eff}^2 + \sigma^2_{\rm clas} \,. \label{eq:clas}
\end{align}
Hence, decoherence due to classical averaging has precisely the same effect as intrinsical quantum mechanical decoherence \cite{Kiers:1995zj,Stodolsky:1998tc,Ohlsson:2000mj}.\footnote{This statement is consistent with the comments related to the integration over $E_P$ after \cref{eq:R_D}.} These results reflect the following two (rather obvious) statements: ($i$) quantum mechanical uncertainties provide a fundamental lower bound on classical uncertainties, and ($ii$) in order to observe effects of quantum mechanical decoherence, classical averaging effects have to be suppressed down to the quantum level.

\section{Numerical estimates}
\label{sec:numerics}

The QFT formalism outlined above allows to calculate the relevant uncertainties $\delta_{\rm loc}$ (or equivalently $\sigma_m$) and $\sigma_{\rm en}$ relevant for the localization and energy spread decoherence factors $\xi_{\rm loc}$ and $\xi_{\rm en}$, respectively, from the properties of the involved external particles. The required input for their definitions in \cref{eq:sigmam,eq:sigma_eff} are the effective energy and momentum uncertainties \cref{eq:sigmap,eq:sigmaE}, which in turn are derived from the momentum spreads of all the external particles in the production and detection processes, as well as their velocities, as defined in \cref{eq:v}. We will now evaluate $\delta_{\rm loc}$ and $\sigma_{\rm en}$ for reactor neutrino and Gallium radioactive source experiments.

\subsection{Particle localizations and velocities}
\label{sec:localisation}

First we need to estimate the momentum
spreads $\sigma$ of all involved particles, as well as their
velocities $v$. Similar estimates have been performed recently in \cite{Akhmedov:2022bjs} in the context of neutrino wave packets. The momentum spread is calculated via the spatial localization $\delta_x$, assuming the uncertainty principle $\delta_x\sigma = 1/2$. We list the relevant quantities for all the particles involved in the production and detection processes in reactor and Gallium experiments in \cref{tab:spreads}. They are estimated as follows.

\begin{table}
  \centering
  \begin{tabular}{l@{\quad}|c@{\quad}|@{\quad}c@{\quad}c@{\quad}c}
    \hline\hline
    & Particle & $\delta_x$ [nm] & $\sigma$ [eV] & $v$ \\
    \hline
    Reactor ($P$) &
    $N$ & 0.24 & 410 & $1\times 10^{-6}$ \\
    $N\to N'+e^-+\overline{\nu}_e$ &
    $N'$ & 0.24 & 410 & $4\times 10^{-5}$ \\
    & $e^-$ & 260 & 0.38 & 0.99 \\
    \hline
    Reactor ($D$) &
    $p$ & 0.1 & 990 & $5\times 10^{-6}$\\
    $p+\overline{\nu}_e \to n + e^+$ &
    $n$ & $5\times 10^6$ & $2\times 10^{-5}$ & $5\times 10^{-3}$ \\
    & $e^+$ & 320 & 0.3 & 0.99 \\
    \hline\hline
    Gallium ($P$) & Cr &0.20 &480&\num{7e-7}\\
    Cr $\to$ V $+\nu_e$ & V &0.20&480&\num{2e-5}\\
    \hline
    Gallium ($D$) & Ga &0.27&370&\num{6e-7}\\
    Ga $+\nu_e \to$ Ge $+e^-$ & Ge &0.27&370 & \num{1e-5}\\
    & $e^-$ & 310 & 0.32 & 0.83\\
    \hline\hline
  \end{tabular}
  \mycaption{Spatial localization $\delta_x$, momentum spread $\sigma = 1/(2\delta_x)$, and velocity $v$ of the external particles involved in the production ($P$) and detection ($D$) processes (first column) of reactor and Gallium source experiments.
 \label{tab:spreads}}  
\end{table}

\paragraph{Reactor experiments.}

For the initial and final state nuclei $N$, $N'$ in the production process, via beta decay within the nuclear fuel, we assume that the localization is determined by a typical interatomic distance \cite{giunti}. We estimate this by using that the lattice parameter of uranium oxide UO$_2$ is $a=5.471\times 10^{-10}$~m and has 4 U and 8 O atoms in one unit cell \cite{LEINDERS2015135}, which gives $\delta_x \simeq a/12^{1/3} \simeq 0.24$~nm. For the initial state nucleus we assume a thermal velocity $v = \sqrt{k_B T / m}$, where the temperature in the nuclear fuel ranges from 700~K at the outer egde to 2000~K in the center \cite{ewing}. This is justified, as the fission products termalize on time scales much faster than their beta decay lifetimes \cite{Akhmedov:2022bjs}. For nuclei with mass numbers in the range of 80 to 160, the velocity of fission products lies in the range of $6.3 \times 10^{-7}$ to $1.5 \times 10^{-6}$. For our estimates a typical value of $v_N \simeq 10^{-6}$ is taken. The recoiling nucleus after beta decay, $N'$, is not in thermal equilibrium and we estimate its velocity using $v = |\mathbf{p}|/E$. Assuming typical neutrino energies $E\simeq 4$~MeV and mass numbers from 80 to 160 we find $v_{N'} \simeq 4\times 10^{-5}$. 

The protons in the detector are typically bound in carbon molecules. We assume a localization of $10^{-10}$~m corresponding to the typical size of the C--H bound length. For the velocity we take thermal velocities at room temperature. Concerning the neutron, after being produced it undergoes scattering in the liquid scintilator. The spatial localization is therefore estimated by the mean collision length $l=1/n\sigma_{\rm int}$, where we use the neutron on CH$_2$ scattering cross section from \cite{nscat}. For a typical kinetic neutron energy of $E_{\rm kin} \simeq 10$~keV a cross section of $\sigma_{\rm int} \approx 50$~b leads to a mean collision length of 4.8~mm. Here, the number density of CH$_2$ was estimated as $n \simeq 1/(3 d^3)$ for an interatomic distance $d \simeq 20$~nm. The velocity is calculated for a neutron with $E_{\rm kin} \simeq 10$~keV. 

For the outgoing electron and positron passing through the medium, either in the nuclear fuel or in the detector, we proceed as follows. We consider the mean rate of energy loss $\langle -dE/dx\rangle$ using the ``Bethe equation'' \cite{Workman:2022ynf}, which we numerically integrate. As localization we take the distance which the particle travels until it deposits one mean excitation energy $I$. This should provide us with a good estimate for the mean free path of the particle.

For the calculation of $\langle -dE/dx\rangle$ we need to know the number density of elementary charges, which can be estimated from the density and composition of the stopping medium. For the reactor fuel we use the above mentioned properties of UO$_2$. As a typical detector material we assume linear alkylbenzene (LAB), $\text{C}_6\text{H}_5\text{C}_{n}\text{H}_{2n+1}$, with $n$ ranging from 10 to 13; for definiteness we will assume $n=12$. We calculate the number density of elementary charges as $138 N_A\rho/m_\text{mol}$, where 138 is the number of elementary charges/electrons in one molecule of LAB, and the density of LAB is taken to be $\rho=\SI{0.859}{\g\per\cm\cubed}$ \cite{JUNO:2015zny}. Furthermore, we need the mean excitation energies $I$. As we consider materials made up of different elements, the calculated mean excitation energies were averaged over the contributions from the different atoms.
For atoms with high elementary charge $Z$ a good approximation is $I=\SI{10}{\eV}\cdot Z$ \cite{CHU197223}. Thus, at the source, this is used for $\text{UO}_2$, where $Z=36$ on average and thus $I=\SI{360}{\eV}$. At the detector we have LAB which we simplify to $\text{CH}_2$. Then, the average excitation energy of one molecule can be approximated to be $I\approx \SI{48}{\eV}$ \cite{CHU197223}.
With these numbers we find that an electron (positron) with an initial kinetic energy of 3~MeV deposits one $I$ energy after traveling 260~nm (320~nm), which we take as the localization at the neutrino production (detection).


\paragraph{Gallium source experiments.}

To estimate the particle localizations at Gallium experiments we take the BEST experiment as an example \cite{Barinov:2021asz,Barinov:2022wfh}. At the source $^{51}$Cr undergoes electron capture to become $^{51}$V and an electron neutrino. Approximately \SI{90}{\percent} of the produced neutrinos have an energy of \SI{750}{\keV} while \SI{10}{\percent} have energies of \SI{430}{\keV}. For the sake of defineteness we consider the \SI{750}{\keV} neutrinos. The temperature of the source is approximately the one of the surrounding gallium, namely \SI{300}{\kelvin} \cite{Barinov:2021asz,Barinov:2022wfh}. Therefore, the thermal velocity of $^{51}$Cr (compound of nucleus + electron to be captured) is $v_\text{Cr}=\sqrt{k_BT/m}= \num{7.3e-7}$. The final $^{51}$V is not thermal, instead its velocity is calculated as $v_\text{V}=p_\text{V}/E_\text{V}=\num{1.5e-5}$.
The localizations of the $^{51}$Cr and $^{51}$V are approximately the same and estimated from the cristal lattice dimensions. They are bound in a bcc lattice with 2 atoms in one unit cell of size $a=\SI{2.88e-10}{\m}$ \cite{hermann}. This leads to a spatial localization of
$\delta_\text{Cr,V}\approx a/\sqrt 2 =\SI{0.20}{\nm}$.

At the detection we estimate the thermal velocity of Ga at \SI{300}{\kelvin} to $v_\text{Ga}=\sqrt{k_BT/m_\text{Ga}}=\num{6.3e-7}$. The velocity of Ge is not thermal, instead $v_\text{Ge}=p_\text{Ge}/E_\text{Ge}=\num{1.15e-5}$. For the electron we obtain $v_{e^-}=p_{e-}/E_{e-}=\num{0.83}$. All momenta are estimated to be of the order of $E_\nu=\SI{750}{\keV}$.
With the density of gallium $\rho\approx \SI{6}{\g\per\cm\cubed}$,
we obtain the number density as
$n=\rho N_A / m_\text{mol}$ and $n^{-1/3}=\SI{2.67e-10}{\m}$, which we take as the localization for both the Ga and Ge atoms.
For the localization of the electron we proceed as described for the reactor case and integrate 
$\langle -dE/dx\rangle$ until the electron deposites one mean excitation energy, which for Ga is $I=\SI{11}{\eV}\cdot Z=\SI{341}{\eV}$~\cite{CHU197223}. This leads to $\delta_e \approx 310$~nm. 

\bigskip

Our estimates for the particles involved in the production processes reported in \cref{tab:spreads} are in agreement with the results of \cite{Akhmedov:2022bjs} within about one order of magnitude. Ref.~\cite{Akhmedov:2022bjs} does not provide detailed estimates for the detector particles. 

\subsection{Effective energy-momentum spreads and decoherence parameters}
\label{sec:eff-spread}

Now we are in the position to calculate the relevant effective uncertainies. For convenience we summarize again the corresponding relations, as derived in \cref{sec:single-vertex,sec:oscAmpl}. Here, $I=P,D$ labels production and detection processes, and the sum is over all external particles $a$ (initial and final state) of the respective processes:
\begin{align}
  &\mathbf{v}_I \equiv \frac{1}{\sigma_{pI}^2} \sum_{a \in I} \sigma_{a}^2 \mathbf{v}_{a} \,,\qquad
   \Sigma_I \equiv \frac{1}{\sigma_{pI}^2} \sum_{a\in I} \sigma_{a}^2 \mathbf{v}_{a}^2 \,, \label{eq:v2}\\
  &\sigma_{pI}^2 \equiv \sum_{a \in I}\sigma^2_{a} \,,\qquad\qquad
   \sigma_{eI}^2 \equiv \sigma_{pI}^2(\Sigma_I - \mathbf{v}_I^2) \,, \\
  &\Gamma_{I,\rm col}^2 \equiv \sum_{a\in I} \sigma_{a}^2 \mathbf{v}_{a}^2 = \sigma_{pI}^2 \Sigma_I \,,\\
  &\sigma_{EI}^2 \equiv \sigma_{eI}^2 + \Gamma_{I,\rm col}^2 + \Gamma_{I,\rm dec}^2 = \sigma_{pI}^2(2\Sigma_I - \mathbf{v}^2_I)  + \Gamma_{I,\rm dec}^2 \,. \label{eq:sigmaE2}
\end{align}

\begin{table}
  \centering
  \begin{tabular}{c@{\quad}|@{\quad}c@{\quad}c|c@{\quad}c}
    \hline\hline
    & \multicolumn{2}{c|}{Reactor}
    & \multicolumn{2}{c}{Gallium} \\
    \hline
    & $I=P$ & $I=D$ & $I=P$ & $I=D$ \\
    \hline
    $\tilde{v}_I$ & $2.1\times 10^{-5}$ & $5.3\times 10^{-6}$ & \num{8.2e-6} & \num{6.4e-6}\\
    $\sqrt{\Sigma_I}$ & $6.4\times 10^{-4}$ & $3.1\times 10^{-4}$ & \num{1.1e-5} & \num{5.1e-4}\\
    $\sigma_{pI}$ & 580~eV & 990~eV & 680 eV & 520 eV\\
    $\sigma_{eI}$ & 0.37~eV & 0.30~eV & 0.005 eV & 0.27 eV\\
    $\Gamma_{I,\rm col}$ & 0.37~eV & 0.30~eV  & 0.0076 eV & 0.27 eV \\
    $\sigma_{EI} \approx \sigma_{I,\rm eff}$ & 0.53~eV & 0.43~eV & 0.0092 eV & 0.375 eV\\
    \hline
    $\sigma_{m}$ &  \multicolumn{2}{c|}{500 eV}  & \multicolumn{2}{c}{390 eV} \\
    $\delta_{\rm loc}$ &  \multicolumn{2}{c|}{0.20 nm}  & \multicolumn{2}{c}{0.25 nm} \\
    $\sigma_{\rm eff}$ &  \multicolumn{2}{c|}{0.33 eV}  & \multicolumn{2}{c}{0.0092 eV} \\
    $\sigma_{\rm en}$ &  \multicolumn{2}{c|}{0.43 eV}  & \multicolumn{2}{c}{0.5 eV} \\
    $\delta_{\rm en}$ &  \multicolumn{2}{c|}{230 nm}  & \multicolumn{2}{c}{200 nm} \\    
    \hline\hline
  \end{tabular}
  \mycaption{Effective velocities and energy-momentum spreads at neutrino production ($P$) and detection ($D$) for reactor and Gallium source experiments. In the lower part we give the spreads relevant for the energy and localization decoherence terms $\xi_{\rm en}$ and $\xi_{\rm loc}$.
For convenience we give values for $\sigma$'s in eV and $\delta$'s in nm; in natural units they are related by $\sigma_{\rm en} \delta_{\rm en} = 1/2$ and $\sigma_{m} \delta_{\rm loc} = 1/2$.  
    \label{tab:eff_uncert}}
\end{table}

The upper part of \cref{tab:eff_uncert} shows our results for these quantities based on the input from \cref{tab:spreads}.
Momentum spreads $\sigma_{pI}$ are dominated in all cases by the localization of hadronic particles and are of order few 100~eV to keV.
Some comments are in order concerning the velocity. The velocity vector $\mathbf{v}_I$ defined in \cref{eq:v2} depends on the relative orientation of the individual particle velocities $\mathbf{v}_a$. The quantity $\tilde{v}_I$ given in \cref{tab:eff_uncert} is calculated by
\begin{align}
  \tilde{v}_I \equiv \frac{1}{\sigma_{pI}^2} \sum_{a \in I} \sigma_{a}^2 |\mathbf{v}_{a}| \ge |\mathbf{v}_I| \,.
\end{align}
Hence, we obtain an upper bound on $|\mathbf{v}_I|$.
From the table we see that effective velocities and velocity spreads are all $\ll 1$, which implies $\sigma_{eI} \ll \sigma_{pI}$, and the $\sigma_{eI}$ are approximately given by the momentum spreads of the final state leptons $e^\pm$, except for the electron capture production (which has no final state charged lepton). Furthermore, $|\mathbf{v}_I| \le \tilde{v}_I \ll \sqrt{\Sigma_I}$ implies that $\sigma_{eI} \approx \Gamma_{I,\rm col}$. Again, an exception is the production in Gallium experiments, where the approximation $\tilde{v}_I \ll \sqrt{\Sigma_I}$ is not very good; nevertheless, $\Gamma_{P,\rm col}$ is still of the same order as $\sigma_{eP}$ also in this case.
Finally, for typical lifetimes for fission products of order 1~ms or larger, decay widths are at most of order $\Gamma_{P,\rm dec} \sim 10^{-12}$~eV, and the decay width of $^{51}$Cr is $\Gamma_{\rm Cr,dec} \approx 4\times 10^{-22}$~eV. Hence, the decay widths are always completely negligible compared to $\sigma_{eI}$ and $\Gamma_{I,\rm col}$ and can be savely neglected in \cref{eq:sigmaE2} and we have $\sigma_{EI} \approx \sqrt{2\Sigma_I} \sigma_{pI} \approx \sqrt{2} \sigma_{eI}$. Therefore, also $\sigma_{EP}$ and $\sigma_{ED}$ are determined by the momentum spread (or by the localization) of the outgoing charged lepton (again with the exception of the electron-capture source).

Now we can move to the calculation of the effective uncertainties summarized in the lower part of \cref{tab:eff_uncert}. We start with $\sigma_m$, defined as
\begin{align}
  \frac{1}{\sigma_m^2}  \equiv  \sum_{I=P,D} \left(\frac{1}{\sigma_{pI}^2} + \frac{v_I^2}{\sigma_{EI}^2} \right)\,.
  \label{eq:sigmam2}  
\end{align}
Here, $v_I$ is actually the length of the component of $\mathbf{v}_I$ parallel to the neutrino propagation direction. The small values of $v_I\le\tilde{v}_I$ imply that for reactors, $\sigma_m$ is dominated by $\sigma_{pI}$, whereas in Gallium the small value of $\tilde{v}_P$ is partially compensated by the small value of $\sigma_{EP}$, and the energy term gives a non-negligible contribution to $\sigma_m$, although the order of magnitude remains unchanged and $\sigma_m \sim \sigma_{pI}$. We find for both type of experiments values of order
\begin{align}
  \sigma_m \simeq (400 - 500) \, {\rm eV} \,,\qquad \delta_{\rm loc} = \frac{1}{2\sigma_m} \simeq 0.2 \,{\rm nm}\,,
  \label{eq:loc}
\end{align}
For the localization term $\xi_{\rm loc}$ from \cref{eq:xi_loc} this implies
\begin{align}
  -\ln\xi_{\rm loc} =
   \frac{1}{2} \left(\frac{\Delta m^2}{4E_\nu\sigma_m}\right)^2 
  \approx 1.3 \times 10^{-19}
  \left(\frac{\Delta m^2}{1\,\rm eV^2}\right)^2
  \left(\frac{1\,\rm MeV}{E_\nu}\right)^2
  \left(\frac{500\,\rm eV}{\sigma_m}\right)^2 \,.
\end{align}
Hence, $\xi_{\rm loc} = 1$  for all practical purposes and localization decoherence is irrelevant in reactor and Gallium experiments. We emphasize, however, that classical spatial averaging in real experiments can be significant, with $\delta_{\rm clas} \sim 1$~m (corresponding to typical sizes of reactor cores or Gallium detectors), which, depending on the value of $\Delta m^2$, does play an important role and has to be included in the analyis of these types of experiments.

Moving to the energy spread, we note that
\begin{align}
  \frac{1}{\sigma_{I,\rm eff}^2} \equiv  \frac{1}{\sigma_{pI}^2} + \frac{(1 - v_I)^2}{\sigma_{EI}^2}
  \approx \frac{1}{\sigma_{EI}^2} \,.  \label{eq:sigma_eff2}  
\end{align}
In the last relation in \cref{eq:sigma_eff2} we have used that $v_I \le \tilde{v}_I \ll 1$ and $\sigma_{EI} \ll \sigma_{pI}$ to obtain $\sigma_{I,\rm eff} \approx \sigma_{EI}$. Hence, for the effective energy spread we obtain
\begin{align}
  \sigma_{\rm eff} \equiv \left[\sum_{I=P,D} \frac{1}{\sigma_{I,\rm eff}^2}\right]^{-1/2}
  \approx \left[\sum_{I=P,D} \frac{1}{\sigma_{EI}^2}\right]^{-1/2}
  \approx
  \left\{
  \begin{array}{l@{\quad}l}
  0.33 \,{\rm eV}  & \text{(reactor)}\,, \\
  0.0092 \,{\rm eV}  & \text{(Gallium)}\,, \\    
  \end{array}\right.
\end{align}
dominated by the outgoing $e^\pm$ momentum spreads for reactors and by the electron capture decay for Gallium experiments, the latter leading to a value of $\sigma_{\rm eff}$ more than one order of magnitude smaller.


As discussed above, $\sigma_{\rm eff}$ corresponds to the ``pure quantum mechanical'' energy spread, c.f., \cref{eq:Asq2}. However, even in idealized experimental configurations there is some un-avoidable averaging, e.g., due to phase-space integrals over unobserved momenta. Therefore, as argued in \cref{sec:phase-space}, the relevant energy spread in reactor experiments is $\sigma_{D,\rm eff}$, due to the averaging of unobserved momenta at the production region. This does not change the qualitative picture, as in our approximation they are of the same order of magnitude. Numerically we find for the spread relevant for $\xi_{\rm en}$ the values:
\begin{align}
  \sigma_{\rm en} \approx \sigma_{D,\rm eff} \approx 0.43\,{\rm eV}\,,\qquad
  \delta_{\rm en} \approx 230\,{\rm nm} \qquad\text{(reactor)} \,.
  \label{eq:en_rea}
\end{align}

For the Gallium source experiments, it is more natural to integrate first over the phase-space of the detector particles, as no momenta are measured in the detector. This would imply $\sigma_{P,\rm eff}$ as the relevant energy spread. Here the effect of the phase-space integration is even less important, as we anyway have $\sigma_{P,\rm eff} \approx \sigma_{\rm eff} \approx 0.0092$~eV or $\delta_{\rm en} \approx \SI{ 11}{\micro\m}$. However, in this case another fundamental (though classical) averaging effect needs to be taken into account, namely the Doppler broadening due to the thermal motions of the source particles~\cite{giunti,Akhmedov:2022bjs}. This leads to an energy smearing with Gaussian shape, with the width set by
\begin{align}
  \sigma_{\rm Doppler} \simeq v_{\rm Cr} E_\nu \approx 0.5\,\rm eV \,.
\end{align}
We can include this effect in a straight forward way, following the discussion in \cref{sec:classical}. We see that $\sigma_{\rm Doppler} \gg \sigma_{P,\rm eff}$ and therefore it dominates the energy spread. Using \cref{eq:clas} with $\sigma_{\rm clas} = \sigma_{\rm Doppler}$, we obtain
\begin{align}
  \sigma_{\rm en} \approx \sigma_{\rm Doppler} \approx 0.5\,{\rm eV}\,,\qquad
  \delta_{\rm en} \approx 200\,{\rm nm} \qquad\text{(Gallium)} \,,
  \label{eq:en_gal}
\end{align}
quite similar to the values obtained for reactor experiments in \cref{eq:en_rea}.

Note that Doppler broadening is in principle relevant for reactor experiments as well. In the neutrino source it would, however, not induce additional decoherence, as we already integrate over the effective neutrino energy in the source and the additional smearing due to the Doppler broadening would have no effect. However, it does contribute to the energy resolution of the detector. Here, we consider it as part of the classical energy resolution and do not include it in the spreads given in \cref{eq:en_rea}.

For the energy spread decoherence $\xi_{\rm en}$ written as in \cref{eq:xi_en2}, we find
\begin{align}
  -\ln\xi_{\rm en} =
   2\pi^2 \left(\frac{L}{L_{\rm osc}}\frac{\sigma_{\rm en}}{E_\nu}\right)^2 
  \approx 4.9\times 10^{-12} \, \left(\frac{L}{L_{\rm osc}}\right)^2
  \left(\frac{1\,\rm MeV}{E_\nu}\right)^2
  \left(\frac{\sigma_{\rm en}}{0.5\,\rm eV}\right)^2 \,, \label{eq:xi_en_num}
\end{align}
which again implies $\xi_{\rm en} = 1$, both for reactor and Gallium experiments to very good accuracy.

The energy resolution of typical detectors in reactor experiments is in the range $(0.03 - 0.06) \, {\rm MeV} \sqrt{E / {\rm MeV}}$ \cite{An:2016ses,PROSPECT:2020sxr,JUNO:2015zny} and hence, $\sigma_{\rm clas} \simeq 0.1$~MeV, about 6 orders of magnitude larger as the intrinsic energy spread \cref{eq:en_rea}. \Cref{eq:xi_en_num} shows that for these values, $-\ln\xi_{\rm en}$ can become of order one, which just means that the (classical) averaging due to the energy resolution of neutrino detectors is an important effect. On the other hand consider the upcoming high precision JUNO reactor experiment~\cite{JUNO:2015zny}, which aims to observe oscillations due to the mass squared difference $\Delta m^2_{31} \simeq 2.5\times 10^{-3}$~eV$^2$ at a distance of about 53~km. This implies $L/L_{\rm osc} \sim 13$ and $-\ln\xi_{\rm en} \sim 5\times 10^{-11}$ for $E_\nu \simeq 4$~MeV. Therefore, quantum decoherence effects will remain completely unobservable also for JUNO.

\paragraph{Comparison with previous results.} Our results of the uncertainties in \cref{eq:en_rea,eq:en_gal} are in rough agreement with \cite{Akhmedov:2022bjs}, but differ by a factor of $10^3-10^4$ from \cite{Jones:2022hme} where $\delta_{\rm en} \simeq 10-400$~pm is obtained. The authors of Ref.~\cite{Jones:2022hme} consider localization scales induced by nucleon-nucleon interactions within the nucleus as well as the size of the decaying nucleon. However, we argue that dynamics at length scales much smaller than $1/|q|$, with $q$ being the typical momentum transfer of the reaction, are irrelevant to the problem, as localizations smaller than $1/|q|$ cannot be resolved. For the same reason scattering of neutrinos with MeV energies can be considered with the nucleus as a whole, and not on individual nucleons (unless they are effectivley ``free'', as the proton in Hydrogen), or even quarks. Therefore, it is the localization of the nuclues, which is relevant for defining the quantum mechanical uncertainties for the process. See also Ref. \cite{Akhmedov:2022mal}.

The authors of Refs.~\cite{deGouvea:2020hfl,deGouvea:2021uvg} consider a term similar to $\xi_{\rm en}$ and perform a phenomenological analysis using reactor data to set a lower bound on a ``coherence length'', finding $\delta > 2.1\times 10^{-4}$~nm, which corresponds to $\sigma = 1/(2\delta) < 0.47$~MeV. This value has also been adopted by \cite{Arguelles:2022bvt} to study the impact on sterile neutrino oscillations in short-baseline experiments. This value of the decoherence parameter is of similar order as the energy resolution of the detectors, 5--6 orders of magnitude larger than the intrinsic quantum uncertainty, and therefore fully dominated by the classical averaging effect \cite{Akhmedov:2022bjs}. 

\section{Summary}
\label{sec:conclusions}

We have used the QFT approach to neutrino oscillations to estimate quantum decoherence effects for reactor neutrino and Gallium source experiments. In this formalism possible decoherence effects on the oscillation probability are fully determined by specifying the momentum spreads (or equivalently the localizations) of the external particles in neutrino production and detection processes. The neutrino is treated as an internal particle, which is integrated out. It is not necessary to introduce the concept of neutrino wave packets, however, the results can also naturally be interpreted in terms of neutrino wave packets \cite{Akhmedov:2010ms}. Throughout our calculations we adopt the approximation of a Gaussian shape for all functions describing momentum and energy spread. 

In \cref{sec:single-vertex} and \cref{sec:oscAmpl} we have reviewed the QFT formalism and derived the coherence factors in terms of external particle localizations and velocities. We introduced the concept of coherence loss due to frequent interactions with the environment, and we find that for the cases of interest this effect is of a similar size as the finite interaction time due to wave packet overlap of the particles involved in the production and detection processes. Furthermore, in our approach we focus on experimentally observable quantities, such as the positron energy after inverse beta decay in the detector, which is used to reconstruct the neutrino energy. We show that phase-space integrals over unobserved particle momenta lead to additional decoherence effects, which correspond to classical averaging of the amplitude squared, but nevertheless are intrinsic to the experimental configuration and in principle unavoidable. Finally, we recover the well-known result that quantum and classical decoherence effects are experimentally indistinguishable, and phenomenologically this distinction is unphysical. Quantum decoherence can only be observed if all classical effects leading to energy and localization averaging can be suppressed down to the quantum level, which of course is not possible in real-life neutrino experiments.

In \cref{sec:numerics} we have performed numerical estimates of decoherence effects for reactor and Gallium radioactive source experiments. First we estimated the localizations and velocities of all involved particles in the production and detection processes. From these we calculate the effective decoherence parameters relevant for the two types of decoherence related to localization and energy spread, \cref{eq:xi_loc2,eq:xi_en2}, respectively. The main results are summarized in \cref{sec:eff-spread} and \cref{tab:eff_uncert}. Localization decoherence is controlled by the factor $(\delta_{\rm loc}/L_{\rm osc})^2$. With typical values of $\delta_{\rm loc} \sim 0.2$~nm, this is many orders of magnitude smaller than any relevant oscillation length $L_{\rm osc}$ and therefore, localization decoherence is completely negligible for practical purposes. The second decoherence effect is related to the energy spread and around the first oscillation maximum it is controled by $(\sigma_{\rm en}/E_\nu)^2$. In the wave packet picture this term can be interpreted as decoherence due to wave packet separation. Both, for reactor and Gallium experiments we find values of order $\sigma_{\rm en} \sim 0.5$~eV, about 6 to 7 orders of magnitude smaller than typical neutrino energies $E_\nu$, leading again to completely negligible decoherence effects.

Hence, in both cases (localization and energy spread) quantum decoherence effects are completely unobservable and many orders of magnitude smaller than classical averaging effects due to finite neutrino source and detector sizes as well as classical energy resolution effects. This remains true even for the upcoming high precision JUNO reactor experiment. If at some point data might provide evidence for decoherence effects beyond classical averaging, standard quantum mechanical uncertainties as discussed here cannot be responsible and this would definitly require exotic new physics \cite{Barenboim:2006xt,Farzan:2008zv,Bakhti:2015dca,Banks:2022gwq}.

\subsection*{Acknowledgement}

We thank E.~Akhmedov for discussions and W.~Grimus for useful communication about his work~\cite{grimus} on a related topic. This work has been supported by the European Union’s Framework Programme for Research and Innovation Horizon 2020 under grant H2020-MSCA-ITN-2019/860881-HIDDeN.

\appendix

\section{Decoherence integral}
\label{app:int}

The integrals which lead to decoherence terms are all of the type
\begin{align}
  I = \int dx \, h(x) e^{i\phi(x) - g(x)} \,,
\end{align}
where for our case of interest the complex phase is given by $\phi = \Delta m^2 L/(2E)$ and $x$ can be either $L$ or $E$.
The real function $g(x)$ has a minimum at $x=x_0$ and we expand $g$ and $\phi$ up to leading non-trivial order around that minimum:
\begin{align}
  g(x) &\approx g(x_0) + \frac{1}{2}g''(x_0)(x-x_0)^2 \,,\\
  \phi(x) &\approx \phi(x_0) + \phi'(x_0)(x-x_0) \,. \label{eq:exp_phi}
\end{align}
The function $h(x)$ is assumed to be sufficiently smooth on scales
$1/\sqrt{g''(x_0)}$ such that we can approximate it by $h(x_0)$ and
pull it out of the integral. Then, the integral becomes Gaussian and we obtain
\begin{align}
  I \approx h(x_0) e^{i\phi(x_0) - g(x_0)} \sqrt{\frac{2\pi}{g''(x_0)}}
  \exp\left[-\frac{(\phi')^2}{2 g''}\right]_{x_0} \,. \label{eq:int-app}
\end{align}
In our applications the last exponential in this expression leads to decoherence if the argument becomes sizeable.

Keeping quadratic terms for $\phi$ in \cref{eq:exp_phi} would lead to a small correction to the phase $\phi(x_0)$ \cite{Cheng:2022lys}, which is neglected here, as we are interested in the
situation where $\phi(x_0) \sim 1$, i.e., around the first (or maybe second) oscillation
maximum.

\section{Decay rate and scattering cross section}
\label{app:rates}

We depart from the amplitude \cref{eq:A2} derived in \cref{sec:single-vertex} for a single-vertex process, considering the effective energy spread $\sigma_{E}$ from \cref{eq:sigmaE}. 
From this result we obtain the total transition probability as:
\begin{align}\label{eq:P}
  P_{if} = |i\mathcal{A}|^2 = |\mathcal{M}|^2 \prod_{i,f}\mathcal{N}_i^2\mathcal{N}_f^2 
   \frac{\pi^4}{(\sigma_p^3\sigma_E)^2}
  \exp\left[ -\frac{(\Delta\mathbf{p})^2}{2\sigma_p^2}
    -\frac{(\Delta E - \Delta\mathbf{pv})^2}{2\sigma_E^2}\right] \,.
\end{align}
Now we follow Weinberg \cite{weinbergQFT} and outline how to derive decay rates or cross sections from this expression, taking into account the wave packet localization of all external particles. Weinberg considers a cubic box around the interaction point in which all particles are localized. This leads to a discretization of all momenta. Here, we adapt the argumentation to the case when particles are localized as Gaussian wave packets, allowing for different degrees of localization for each particle, see also \cite{Ishikawa:2018koj,Ishikawa:2020hph}.

For each particle $a$ we can consider $V_a \sim 1/\sigma_a^3$ as the volume in which the particle is confined due to its wave packet. It turns out that a suitable definition is
\begin{align}\label{eq:V}
  V_a = \left(\frac{\sqrt{2\pi}}{2\sigma_a}\right)^3 \,.
\end{align}
The numerical factor can be motivated as follows: define the configuration space spread $\delta_x$ by $\delta_x\sigma = 1/2$. Then $V = (\sqrt{2\pi} \delta_x)^3$ and $1/V$ would be just the prefactor of a Gaussian in configuration space with width $\delta_x$. 

Let us now relate the transition probability in \cref{eq:P} to more familiar quantities. Consider first the final state particles. $P_{if}$ is the total transition probability for $i\to f$ where all particles are defined via the respective wave packets. Usually we are interested in a differential quantity, for a transition into an infinitesimal phase space element for the outgoing particles. Such a quantity can be obtained by multiplying the probability by the differential number $dN_f$ of outgoing particles within the phase-space element $d^3p_f$ around momentum $\mathbf{p}_f$:
\begin{align}\label{eq:dP}
  dP_{if} = P_{if} \prod_f dN_f \,,
\end{align}
with
\begin{align}
  dN_f = V_f \frac{d^3p_f}{(2\pi)^3} = \frac{d^3p_f}{(2\sigma_f)^3(2\pi)^{3/2}} \,,
\end{align}
where in the second step we have used \cref{eq:V}.
Hence, each final state factor in \cref{eq:P} becomes just the usual phase-space factor:
\begin{align}
  \mathcal{N}_f^2 dN_f = \frac{1}{2E_{p_f}}\frac{d^3p_f}{(2\pi)^3} = d\tilde{\mathbf{p}}_f \,,
\end{align}
where we have used the definition of $\mathcal{N}$ from \cref{eq:N}.

Next we consider the following factors from \cref{eq:P}:
\begin{align}
  \frac{\pi^4}{(\sigma_p^3\sigma_E)^2} e^{-[\ldots]}
  & = V_I T_I \, (2\pi)^4
  \frac{1}{(\sqrt{2\pi})^4 \sigma_p^3\sigma_E}
    \exp\left[ -\frac{(\Delta\mathbf{p})^2}{2\sigma_p^2}
    -\frac{(\Delta E - \Delta\mathbf{pv})^2}{2\sigma_E^2}\right] \\
  & = V_I T_I \, (2\pi)^4
    \delta^{(3)}_G(\Delta\mathbf{p})
    \delta_G(\Delta E - \Delta\mathbf{pv}) \label{eq:deltaG}\\
  & \to V_I T_I \, (2\pi)^4
    \delta^{(4)}\left(\sum_f p_f - \sum_i p_i\right) \label{eq:deltaGG}
\end{align}
where
we have defined the effective interaction volume and interaction time in analogy to \cref{eq:V} by using the effective momentum and energy spreads from \cref{eq:sigmap,eq:sigmaE}:
\begin{align}
  V_I = \left(\frac{\sqrt{2\pi}}{2\sigma_p}\right)^3 \,,\qquad
  T_I = \frac{\sqrt{2\pi}}{2\sigma_E} \,. \label{eq:VI_TI}
\end{align}
In \cref{eq:deltaG} we have introduced the notation $\delta_G$ to
denote a Gaussian, which will converge to a delta function in the
limit $\sigma\to 0$ and the limit in \cref{eq:deltaGG} is obtained for
$\sigma_p \to 0$.\footnote{We use here $\delta^{(3)}(\Delta\mathbf{p})
    \delta(\Delta E - \Delta\mathbf{pv}) = \delta^{(3)}(\Delta\mathbf{p})
    \delta(\Delta E) = \delta^{(4)}(\Delta p)$.}
Finally, we use \cref{eq:V} also to rewrite the initial state factors
$\mathcal{N}_i$ in terms of volumes and obtain
\begin{align}\label{eq:dGamma}
  d\Gamma_{if} \equiv 
  \frac{dP_{if}}{T_I} = V_I \prod_i \left(\frac{1}{2E_i}\frac{1}{V_i}\right)
  \prod_f d\tilde{\mathbf{p}}_f \, 
(2\pi)^4 \delta^{(3)}_G(\Delta\mathbf{p})
    \delta_G(\Delta E - \Delta\mathbf{pv}) |\mathcal{M}|^2 \,.
\end{align}
We interpret $T_I$ as the effective interaction time. Therefore, we obtain the differential interaction rate $d\Gamma_{if}$ by dividing the differential transition probability by the interaction time.
\Cref{eq:dGamma} is sometimes called Fermi's Golden Rule.

\subsection{Scattering cross section}

Consider now the case of the scattering of two initial state particles to a set of final state particles $AB \to f$. We get the factor $V_I/V_AV_B = V_I n_A n_B$  in \cref{eq:dGamma}, with $n_A, n_B$ denoting the number density of particles $A,B$. Let us consider particle $A$ as ``target''. Then $V_In_A$ is the number of target particles inside the interaction volume, and the flux of incoming $B$ particles is $v_{\rm rel} n_B$, with $v_{\rm rel}$ being the relative velocity. The cross section per target particle is given by (total interaction rate) / [(number of target particles $A$) $\times$ (flux of incoming particles $B$)]. Hence,
\begin{align}
  d\sigma = \frac{d\Gamma_{if}}{V_I n_A v_{\rm rel} n_B} = 
  \frac{1}{2E_A2E_Bv_{\rm rel}} \prod_f d\tilde{\mathbf{p}}_f \, 
(2\pi)^4 \delta^{(3)}_G(\Delta\mathbf{p})
    \delta_G(\Delta E - \Delta\mathbf{pv}) |\mathcal{M}|^2 \,,  
\end{align}
which, in the limit $\sigma_p, \sigma_E\to 0$, where the Gaussian $\delta_G$ functions become true $\delta$-functions, converges to the standard expression for a cross section, e.g., \cite{peskin}.

\subsection{Particle decay}
\label{app:decay}

Consider now the case of a single particle in the initial state, i.e., the decay of a particle. 
In this case we obtain a factor $V_I/V_i$ in \cref{eq:dGamma}. We can consider $n_i = 1/V_i$ as the density of initial state particles and $V_I$  as the effective interaction volume, with $V_I \le V_i$. Hence, $V_I n_i$ is the number of initial-state particles inside the interaction volume. Usually we are interested in the decay rate of a single particle, denoted just by $\Gamma_{\rm dec}$.
Therefore, we obtain
\begin{align}
  d\Gamma_{\rm dec} \equiv \frac{d\Gamma_{if}}{V_I n_i} =
  \frac{1}{2E_i} \prod_f d\tilde{\mathbf{p}}_f \, 
(2\pi)^4 \delta^{(3)}_G(\Delta\mathbf{p})
    \delta_G(\Delta E - \Delta\mathbf{pv}) |\mathcal{M}|^2 \,.  
\end{align}
In the limit $\sigma_p\to 0$, where the Gaussian $\delta_G$ functions become true delta-functions, 
this expression converges to the standard expression for the decay rate of a particle, see e.g., \cite{peskin}.

Let us now comment on the considerations in \cref{sec:interact}
related to collisions with the environment. We restore the Lorentzian
shape of the energy spread, see \cref{eq:lorentzian}, and instead of \cref{eq:dGamma} we obtain
\begin{align}
  d\Gamma_{if} \equiv 
  \frac{dP_{if}}{T_I} = V_I \prod_i \left(\frac{1}{2E_i}\frac{1}{V_i}\right)
  \prod_f d\tilde{\mathbf{p}}_f \, 
(2\pi)^3 \delta^{(3)}_G(\Delta\mathbf{p})
  |\mathcal{M}|^2 \,\frac{\Gamma}{\Omega^2 + \Gamma^2/4} \,, \label{eq:dG_app}
\end{align}
where now we identify $T_I = 1/\Gamma$. We recognise that the energy conservation function $\delta_G(\Omega)$ is now replaced by the Lorentzian, whose width is set by the rate $\Gamma$. We consider now two physically different situations:
\begin{itemize}
\item 
First we assume that the collision rate $\Gamma_{\rm col}$ is much larger than the decay rate $\Gamma_{\rm dec}$ and set $\Gamma = \Gamma_{\rm col}$ in \cref{eq:dG_app}.
Let us consider as simple example a massive scalar particle (mass $M_i$) at rest decaying into two massless scalar particles. The matrix element is a constant, and we can perform the phase space integrals, apart from the energy of one of the final state particles and obtain
\begin{align}
  \frac{d\Gamma_{\rm dec}}{dE_f} = \frac{|\mathcal{M}|^2}{16\pi^2M_i}
  \frac{\Gamma_{\rm col}}{(M_i - 2E_f)^2 + \Gamma_{\rm col}^2/4} \,.
\end{align}
We observe the collisional line broadening \cite{VanVleck:1945zz}, i.e., the width of the
decay line is set by the collision rate. Integrating over $E_f$ leads to the same total decay rate as for $\Gamma_{\rm col}\to0$, namely $\Gamma_{\rm dec} = |\mathcal{M}|^2/(16\pi M_i)$.

\item
  Let us now assume that $\Gamma_{\rm dec} \gg \Gamma_{\rm col}$, i.e., we can neglect collisions. 
  Now we can set $\Gamma = \Gamma_{\rm dec}$ in \cref{eq:dG_app} and use that the probability that the particle decays into any final state is one: $\int dP_{if} = 1$. Considering again the
example of a massive scalar particle at rest decaying into two massless scalar particles, we obtain from 
\cref{eq:dG_app} in the limit of exact momentum conservation $\delta^{(3)}_G(\Delta\mathbf{p}) \to \delta^{(3)}(\Delta\mathbf{p})$:
\begin{align}
  1 = \frac{|\mathcal{M}|^2}{8\pi^2 M_i} \frac{\arctan(2M_i/\Gamma_{\rm dec})}{\Gamma_{\rm dec}} \,.
\end{align}
In the limit $\Gamma_{\rm dec} \ll M_i$ we can solve for $\Gamma_{\rm dec}$ and recover the standard result for the decay width $\Gamma_{\rm dec} = |\mathcal{M}|^2/(16\pi M_i)$.
\end{itemize}

\subsection{Rate for the neutrino oscillation process}
\label{app:rate-osc}

In the case of neutrino oscillations, the observable of interest is the differential event rate at the detector $dR_D$, depending on the momenta of the final states in the detector, i.e., $\mathbf{p}_3,\mathbf{p}_4$ in our notation. In the case of reactor neutrino experiments these are the positron and neutron, whose momenta are in principle observable in the detector. We can obtain this quantity from the transition probability $P_{if} = |i\mathcal{A}_{\alpha\beta}|^2$ by the following steps:
\begin{enumerate}
\item As in \cref{eq:dP}, we multiply by the phase-space elements of all final state particles in order to obtain a differential transition probability,
  \begin{equation}
  dP_{if} = P_{if} \prod_f dN_f = P_{if} dN_1dN_2dN_3dN_4 \,.      
  \end{equation}
\item We are interested in the event rate per single detector particle, as well as single decaying particle at the source. Hence, we have to divide by the number of particles in the production and detection regions, $V_P/V_A$ and $V_D/V_B$, respectively, with the volumina defined as in \cref{eq:V,eq:VI_TI}.
\item The outgoing particles in the production region are not observed. Hence we have to integrate over their phase space
  $d\tilde{\mathbf{p}}_1d\tilde{\mathbf{p}}_2$.
\item
  The event rate at the detector is obtained by dividing by the effective detection time interval $T_D$ defined as in \cref{eq:VI_TI}, in analogy to the first relation in \cref{eq:dGamma}. 
\item
  The amplitude squared still depends on production and detection times via $T = t_D - t_P$. While the detection time in principle is observable in real-time neutrino detectors, the production time typically is not observable (see comments below). Hence, for a given $t_D$, we have to sum the contributions of all possible production times contributing to the amplitude. This is obtained by first dividing by the effective production time interval $T_P$ defined as in \cref{eq:VI_TI}, which gives the neutrino production rate, which then has to be integrated over all possible values of $t_P$. By a simple coordinate shift, this integral is transformed into an integral over $T$ and we obtain a time averaged amplitude-squared:
  \begin{equation}\label{eq:Asq-average}
    \overline{|\mathcal{A}_{\alpha\beta}|^2} = \int dT \,\frac{|\mathcal{A}_{\alpha\beta}(T)|^2}{T_P} \,. 
  \end{equation}
\end{enumerate}
Putting everything together, we obtain for the differential event rate at the detector
\begin{equation}\label{eq:event_rate}
  dR_D = \frac{V_P}{V_A} \frac{V_D}{V_B} \, dN_3dN_4 \int dN_1dN_2 \frac{\overline{|\mathcal{A}_{\alpha\beta}|^2}}{T_D} \,.
\end{equation}

Following comments are in order:
\begin{itemize}
\item In reactor experiments typically only the energy of the outgoing positron is used to reconstruct the neutrino energy. In this case we need also to integrate over the phase space of the outgoing neutron in the detector.
\item In counting experiments, such as Gallium radioactive source experiments, the detection reaction is not observed and we have to integrate in addition over the phase-space of outgoing detector particles $dN_3dN_4$.  
\item
  \Cref{eq:event_rate} corresponds to the hypothetical ``event rate'' assuming a single decaying particle at production and a single detector particle. For an actual experiment we have to sum over the number of target particles in the detector, as well as the total number of all decays contributing to the neutrino production at the source. The latter is a complicated calculation in case of reactor neutrinos, amounting to an ab initio calculation of the total reactor flux.
\item Above we assumed that the neutrino production time is not observable. This is true for a large number of experiments, such as reactor, solar, or atmospheric neutrinos. At many accelerator experiments, neutrinos are produced by a pulsed particle beam, which in principle provides some time information on the production. Typically the duration of the beam spills are of order $\si{\micro\s}$, which however, in practice is much longer compared to other relevant time scales $\sim 1/\sigma_{EP}, 1/\sigma_{ED}$. Therefore, for practical purposes it is a good approximation to assume the integration in 
  \cref{eq:Asq-average} over an infinite time interval \cite{Grimus:2019hlq}.
\item
  In the case of a so-called monitored neutrino beam, such as proposed by ENUBET~\cite{Longhin:2022tkk}, in principle time information on neutrino production can be obtained. In such a situation the amplitude needs to be averaged over the accuracies with which both $t_D$ and $t_P$ can be determined, instead of an infinite time interval. In order to affect the discussion above, the accuracy of both time measurments, $t_D$ and $t_P$, need to be comparable to the microscopic time scales  
$1/\sigma_{EP}$ and $1/\sigma_{ED}$.
\item
  Using the formalism discussed in this paper, one can show \cite{Raphael-MSc} that in the (realistic) case, when decoherence effects play no role, \cref{eq:event_rate} factorizes into the decay rate $\Gamma_P$ of the production process, the standard oscillation probability and the detection cross section $\sigma_D$ as
  \begin{equation}
    dR_D =  \frac{1}{4\pi L^2} \int dE_\nu \frac{d\Gamma_P}{dE_\nu} \, P_{\alpha\beta}(E_\nu,L) \, d\sigma_D (E_\nu) \,,
  \end{equation}
see also \cite{Cardall:1999ze,Kovalenko:2022goz,grimus} for similar calculations.
\end{itemize}

\section{Phase-space integrals}
\label{app:phase-space}

As discussed in \cref{sec:phase-space}, depending on the physical configuration, integrals over the momenta of unobserved external particles either in the source or in the detector are unavoidable, which leads to integrals over the effective neutrino energy at production $E_P$ or at detection $E_D$ (or both). Here we outline the integral of \cref{eq:Asq2} over $E_P$;
the integration over $E_D$ instead of $E_P$ proceeds in complete analogy.

We expand the oscillation phase around $E_D$ and keep terms up to linear order in
$(E_P-E_D)$, using $E_0-E_D = (\sigma_{\rm eff}^2/\sigma_{P,\rm eff}^2)(E_P-E_D)$.
But we replace $E_0 \to E_D$ in the two decoherence terms
in the second line of \cref{eq:Asq2}, ignoring higher order corrections to these terms.
Then the integral is of the form discussed in \cref{app:int} and we obtain
\begin{align}
  \int dE_P  & \exp\left[i\frac{\Delta m^2L}{2E_0}\right] \times
  \exp\left[ -\frac{1}{2} \frac{(E_D-E_P)^2}{\sigma_{P,\rm eff}^2+\sigma_{D,\rm eff}^2} \right] \nonumber\\
  & \propto
  \exp\left[i\frac{\Delta m^2L}{2E_D}\right] \times
  \exp\left[-\frac{1}{2} \left(\frac{\Delta m^2 L}{2E_D^2}\right)^2
    \frac{\sigma_{D,\rm eff}^4}{\sigma_{P,\rm eff}^2+\sigma_{D,\rm eff}^2}\right] \,.
\end{align}
The emerging decoherence term can be combined with the $\sigma_{\rm eff}$-term in \cref{eq:Asq2}.
Using the definition of $\sigma_{\rm eff}$ in \cref{eq:E0} we obtain:
\begin{align}
  R_D(L,E_D) \propto \int dE_P \overline{|\mathcal{A}_{\alpha\beta}|^2} & \propto
  \exp\left[i\frac{\Delta m^2L}{2E_D}\right] \times
  \exp\left[ - \frac{\mathbf{p}_{D\perp}^2 }{2\sigma_{pD}^2} \right]  
  \nonumber\\ 
  &\times \exp\left[
    -\frac{1}{2} \left(\frac{\Delta m^2}{4E_D\sigma_m}\right)^2
    -\frac{1}{2} \left(\frac{\Delta m^2 L}{2E_D} \frac{\sigma_{D,\rm eff}}{E_D}\right)^2
    \right] \,. \label{eq:Asq3}
\end{align}

We have dropped the term depending on $\mathbf{p}_{P\perp}$ in \cref{eq:Asq3}, as the oscillation phase no longer depends on it, and therefore the integration over
$\mathbf{p}_{P\perp}$ becomes trivial. In contrast, $E_D$ -- and therefore the oscillation phase -- does depend on $\mathbf{p}_{D\perp}$. In principle, $\mathbf{p}_{D\perp}$ is observable, by reconstructing all particles involved in the detection process. However, this is often not the case in realistic situations. For instance, in reactor neutrino experiments, the momentum of the outgoing neutron cannot be observed. Therefore, we have to integrate also over the neutron momentum phase-space, which effectively means integrating over $\mathbf{p}_{D\perp}$. Once again we can involve \cref{app:int} to derive a corresponding decoherence term. There is no simple closed form for this term, however, its size can be estimated to be of order $\mathbf{v}_D^2 \sigma_{pD}^2 / \sigma_{D,\rm eff}^2$ relative to the last term in \cref{eq:Asq3}. One can show that $\mathbf{v}_D^2 \sigma_{pD}^2 / \sigma_{D,\rm eff}^2 < 1$ and in typical cases it is actually $\ll 1$. Using the numbers from \cref{sec:numerics} we obtain for reactor neutrinos  $\mathbf{v}_D^2 \sigma_{pD}^2 / \sigma_{D,\rm eff}^2 \sim 10^{-4}$. Hence this term can at most induce corrections of order one to the last term in \cref{eq:Asq3}.

\bibliographystyle{JHEP_improved}
\bibliography{./refs}

\end{document}